\documentclass[aps,prl,twocolumn,floatfix,nofootinbib,showpacs,superscriptaddress]{revtex4}

\pdfoutput=1
\usepackage{amsmath,amsfonts,amssymb,bm,ulem}

\usepackage{graphicx}
\usepackage{color}
\usepackage{url} 
\usepackage{threeparttable}
\usepackage{makecell}

\definecolor{purple}{rgb}{0.1,0.5,0.4}

\usepackage[
pdfauthor={derajan},
pdfstartview=XYZ,
bookmarks=true,
colorlinks=true,
linkcolor=blue,
urlcolor=blue,
citecolor=blue,
pdftex,
bookmarks=true,
linktocpage=true,   
hyperindex=true
]{hyperref} 

\begin{document}

\title{Clockwork axion footprint on nano-hertz stochastic gravitational wave background}

\author{Bo-Qiang Lu} 
\email{bqlu@zjhu.edu.cn}
\affiliation{School of Science, Huzhou University, Huzhou, Zhejiang 313000, P. R. China}

\author{Cheng-Wei Chiang}
\email{chengwei@phys.ntu.edu.tw}
\affiliation{Department of Physics and Center for Theoretical Physics, National Taiwan University, Taipei 10617, Taiwan}
\affiliation{Physics Division, National Center for Theoretical Sciences, Taipei 10617, Taiwan}

\author{Tianjun Li}
\email{tli@itp.ac.cn}
\affiliation{CAS Key Laboratory of Theoretical Physics, Institute of Theoretical Physics, \\Chinese Academy of Sciences, Beijing 100190, P. R. China}
\affiliation{School of Physical Sciences, University of Chinese Academy of Sciences, No.~19A Yuquan Road, Beijing 100049, P. R. China}

\begin{abstract}

The recent Pulsar Timing Arrays (PTAs) nano-Hz gravitational wave (GW) background signal can be naturally induced by the annihilation of domain walls (DWs) formed at a symmetry-breaking scale $f\simeq 200$~TeV in the clockwork axion framework.
Based on our first successful and precise prediction, we for the first time suggest that the recent PTA observations strongly support the novel mechanism of the QCD instanton-induced DW annihilation in the clockwork axion framework.
We also for the first time discover a novel correlation between dark matter (DM) relic abundance and nano-Hz GW background, which in turn indicates a natural connection between the axion decay constant and the symmetry-breaking scale in the clockwork framework.
We find that the GW signal has a peak $h^2\Omega_{\rm GW}\simeq 10^{-6.6}-10^{-6.1}$ at about 50~nHz, which is definite and testable for future PTA data at frequencies $\gtrsim 25$~nHz and CMB-S4 experiment.
We also propose various phenomena that may appear in PTAs and future GW interferometers.

\end{abstract}
\maketitle


{\it Introduction.---}
It is well known that DWs produced with the spontaneous breaking of a discrete $Z_N$ symmetry could dominate the Universe if they do not completely annihilate at a later time, and thus bury the predictions of standard cosmology~\cite{Zeldovich:1974uw,Sikivie:1982qv}.
The DWs formed above the GUT scales $M_{\rm G}\sim 10^{16}$~GeV might be adequately diluted by the inflation. 
Below $M_{\rm G}$, to break the degenerate vacua, one commonly introduces a bias potential with an energy magnitude $\left(v/M_{\rm pl}\right)^2 v^4\lesssim V_{\rm bias}\lesssim \left(v/M_{\rm pl}\right) v^4$ ($v$ is the spontaneous symmetry-breaking scale, and $M_{\rm pl}$ is the Planck scale), where the lower bound ensures the DWs annihilate before dominating the universe and the upper bound indicates the DWs annihilate immediately after their formation. 
The symmetric part of the potential has an energy $V_{\rm sym}\sim v^4$.
A hierarchy arises between the bias and the symmetric parts of the potential if the symmetry breaks much below the GUT scale.
Since many models do not explicitly break the discrete symmetry, the bias term is often introduced as an extrinsic and free parameter while ignoring the hierarchy.
In fact, the small explicit symmetry-breaking can be dynamically induced by some non-perturbative instanton effects if the global symmetry is anomalous in the underlying theory~\cite{Sikivie:1982qv,Preskill:1991kd,Dine:1993yw,Dvali:1994wv,Abel:1995wk}. Two well-known examples are quantum gravity (QG)~\cite{Lee:1988ge,Abbott:1989jw,Coleman:1989zu} and QCD instanton~\cite{Sikivie:1982qv,Gross:1980br,Schafer:1996wv}.
It is believed that all global symmetries are not respected by the QG effect~\cite{Banks:1988yz,Rai:1992xw,Banks:2010zn,Harlow:2018tng,Witten:2017hdv,Antinucci:2023uzq}, which can lead to the decays of DM and DW~\cite{Lattanzi:2007ux,King:2023ayw,Craig:2020bnv}.
The symmetry breaking by the QG effects is described by some higher-dimensional operators.
However, one can not make a predictive announcement since the size of the symmetry-breaking by the QG effect is not well specified.

In Ref.~\cite{Chiang:2020aui}, two of us were the first to notice that the QCD instanton effect could induce a bias potential for the annihilation of DWs formed at a symmetry-breaking scale $f\simeq 200$~TeV in the clockwork axion framework, leading to a loud GW signal for NANOGrav 12.5-year (NG12) observation.
However, this signal behaves as $\Omega_{\rm GW}(\nu)\propto\nu^{\gamma}$ with $\gamma=3$ and thus was not supported by the NG12 data~\cite{NANOGrav:2020bcs}, which can be fitted by a flat spectrum with an amplitude $\Omega_{\mathrm{GW}}(5.5 \mathrm{nHz}) \in\left(3 \times 10^{-10},~2 \times 10^{-9}\right)$ and an exponent $\gamma \in(-1.5,~0.5)$~\cite{Chiang:2020aui}.
During that time, the cosmic string~\cite{Ellis:2020ena,Blasi:2020mfx,King:2020hyd} and primordial black hole (PBH)~\cite{DeLuca:2020agl,Vaskonen:2020lbd} were commonly considered as they generated a flat GW spectrum that aligned with the NG12 data. By fitting the NANOGrav 15-year (NG15) data~\cite{NANOGrav:2023gor} and IPTA-DR2 data~\cite{Antoniadis:2022pcn}, we find that the recent PTA nano-Hz GW observations show a complete consistency with our prediction in Ref.~\cite{Chiang:2020aui}, both in the amplitude and in the exponent of the GW spectrum (see Fig.~10 in Ref.~\cite{Chiang:2020aui} and Fig.~S3 in the Supplemental Material). 

It is worth mentioning that NANOGrav~\cite{NANOGrav:2023gor,NANOGrav:2023hvm}, together with EPTA~\cite{EPTA:2023fyk}, PPTA~\cite{Reardon:2023gzh}, and CPTA~\cite{Xu:2023wog} have recently presented the first convincing evidence for the Hellings-Downs angular correlation, which strongly supports the existence of nano-Hz stochastic GW background (GWB). Although the astrophysical source from a population of inspiraling supermassive black hole binaries (SMBHBs) does not fit better than new physics to the NG15 data, an understanding of the SMBHBs still remains to be improved~\cite{NANOGrav:2023hvm}.
A series of works on nano-Hz GWs also appeared recently, including 
DWs~\cite{King:2023ayw,Guo:2023hyp,Blasi:2023sej,Du:2023qvj,Bai:2023cqj,Zhang:2023nrs}, 
cosmic string~\cite{Ellis:2023tsl,Wang:2023len,Lazarides:2023ksx}, 
scalar-induced GW~\cite{Vagnozzi:2023lwo,Cai:2023dls,Wang:2023ost,Liu:2023ymk,Yi:2023mbm}, 
PBH~\cite{Inomata:2023zup,Franciolini:2023pbf,Bhaumik:2023wmw}, 
first-order phase transition~\cite{Han:2023olf,Megias:2023kiy,Franciolini:2023wjm,Jiang:2023qbm,Zu:2023olm,Xiao:2023dbb},
as well as model comparisons~\cite{Madge:2023cak,Bian:2023dnv,Wu:2023hsa,Ellis:2023oxs}.

In this {\it Letter}, we for the first time systematically present the hierarchy in potential and suggest 
the mechanism of QCD instanton-induced DW annihilation for PTAs, whose novelty rests in three aspects.
First, compared with other interpretations for PTAs, our scenario is predictive since the QCD phase transition definitely takes place in the early Universe. Second, compared with the QG, the size of the bias potential induced by the QCD instanton effect is quantitatively determined by the QCD confinement scale $\Lambda_{\rm QCD}$. Interestingly, the DWs annihilated at the QCD scale can naturally generate nano-Hz GWs for PTAs.   
Finally, the Peccei-Quinn (PQ)  scale is restricted to be much higher than the electroweak scale by the astroparticle physics experiments, which therefore leads to a hierarchy in the potential.
Notice that since the QCD instanton explicitly breaks the $U(1)$ symmetry down to the discrete shift symmetry, the QCD instanton will lead to the formation, rather than the annihilation of the DWs in the classical QCD axion model~\cite{Sikivie:1982qv}.

Although the existence of DM has been confirmed by various cosmological and astrophysical observations~\cite{Planck:2015fie}, the nature of DM still remains unknown. 
Since the bias potential is hierarchically suppressed, the global symmetry should be exact enough to provide a DM candidate. 
This motivates us to further explore the DM phenomenon in the clockwork axion framework.
The axion produced at the symmetry-breaking scale $f$ attains a mass at the QCD scale.
For the first time, we find that one can obtain the correct order of the axion decay constant $f_a$ for the axion DM relic abundance from axion oscillation with the estimate $f_a\sim f^2/\Lambda_{\rm QCD}\sim 10^{11}$~GeV, once we adopt $f\simeq 200$~TeV from the PTA observation.
We expect that the DM relic abundance is closely related to the nano-Hz GWB and the clockwork axion framework provides a natural realization.

{\it GWs from clockwork DW annihilation.---}
The clockwork axion framework~\cite{Choi:2014rja,Choi:2015fiu,Kaplan:2015fuy,Higaki:2015jag,Higaki:2016yqk,Higaki:2016jjh,Giudice:2016yja,Farina:2016tgd,Coy:2017yex,Long:2018nsl,Agrawal:2018mkd}, which was first proposed to break the canonical relation between the symmetry-breaking scale and axion decay constant~\cite{Kaplan:2015fuy}, had recently been extended to other fields in its continuum limit~\cite{Giudice:2016yja,Choi:2017ncj,Wood:2023lis}, thus providing a solution to the Higgs naturalness problem.
The large $N+1$ global $U(1)$ symmetries can appear as an accidental consequence of gauge invariance and 5D locality in an extra-dimensional model~\cite{Kaplan:2015fuy,Arkani-Hamed:2001kyx,Arkani-Hamed:2003xts,Choi:2003wr}.
The framework introduces $N+1$ copies of complex scalars $\Phi_j(x)$ with $j=0,1,...,N$ and the following potential
\begin{equation}
    \label{eq:poten1}
    V(\Phi)=\sum_{j=0}^{N}\left(-m^{2}\left|\Phi_{j}\right|^{2}+\lambda\left|\Phi_{j}\right|^{4}\right)-
    \varepsilon \sum_{j=0}^{N-1}\left( \Phi_{j}^{\dagger} \Phi_{j+1}^{3}+ {\rm H.c.} \right),
\end{equation}
where $m^2$, $\lambda$, and $\varepsilon$ have been assumed to be real and universal.  
The first term respects a global $U(1)^{N+1}$ symmetry and is explicitly broken by the $\varepsilon$-dependent terms down to the $N$ shift symmetries and a global $U(1)$ symmetry, which is identified as the PQ symmetry. 
For $N\gtrsim 3$, the stable string-wall network forms when the radial components acquire a vacuum expectation value $\left \langle \Phi_j \right \rangle=f/\sqrt{2}$, resulting in $N$ massive axions $A_i$ and one massless axion $a$
(see the Supplemental Material).

Because the discrete symmetries are anomalous under the QCD gauge symmetry,
the QCD instanton effects generate a bias potential $V_{\rm bias}\sim \Lambda_{\rm QCD}^4$ (with $\Lambda_{\rm QCD}=(332\pm 17)$~MeV~\cite{ParticleDataGroup:2018ovx}) during the QCD phase transition, which lifts the $N$ degenerate vacua and breaks the residual $U(1)$ symmetry. 
The annihilation of DWs is significant when the surface energy becomes comparable to the bias energy, i.e., 
$\sigma H\simeq V_{\rm bias}$, where $\sigma\simeq 8m_{A}f^2$ is the surface tension of the wall and $m_{A}\simeq \varepsilon^{1/2}f$ is the mass of the massive axion. The DWs quickly annihilate one after another at the temperature 
\begin{equation}
\begin{aligned}
\label{eq:Tann}
T_{\rm ann} \simeq & 7.15 \times 10^{-2} \mathrm{GeV} \varepsilon^{-1 / 4}\left(\frac{g_*\left(T_{\rm ann}\right)}{10}\right)^{-1 / 4} \\
& \times\left(\frac{f}{100~\mathrm{TeV}}\right)^{-3/2}\left(\frac{\Lambda_{\rm QCD}}{100~\mathrm{MeV}}\right)^2.
\end{aligned}
\end{equation}
The GW spectrum from the violent DW annihilation is characterized by a peak frequency determined by $\nu_{\rm{peak }}\left(t_{\rm{ann }}\right)\simeq H\left(t_{\rm{ann }}\right)$, where $t_{\rm ann}$ is the cosmic time of DW annihilation. Then the red-shifted peak frequency today is found to be
\begin{equation}
    \begin{aligned}
    \label{eq:vpeak}
    \nu_{\rm peak}(t_0)\simeq & 1.1 \times 10^{-8} \mathrm{~Hz}\left(\frac{g_{*}\left(T_{\mathrm{ann}}\right)}{10}\right)^{1/2}
    \\
    & \times \left(\frac{g_{*s}\left(T_{\mathrm{ann}}\right)}{10}\right)^{-1/3} \left(\frac{T_{\mathrm{ann}}}{0.1~\rm{GeV}}\right),
\end{aligned}
\end{equation}
where $g_{*}$ and $g_{*s}$ are the effective relativistic degrees of freedom (DOF) associated with energy and entropy, respectively.  
Eq.~\eqref{eq:vpeak} shows that the DW annihilation at the QCD scale naturally induces a GWB at nano-Hz frequencies.

The peak GW amplitude produced at the annihilation time is determined by
$\Omega_{\mathrm{GW}}\left(\nu_{\mathrm{peak}}\left(t_{\mathrm{ann}}\right)\right)
    \simeq\left(8 \pi \tilde{\epsilon}_{\mathrm{gw}} G^{2} A^{2} \sigma^{2}\right)/\left(3 H^2(t_{\rm ann})\right)$, with $\tilde{\epsilon}_{\rm{gw}} \simeq 0.7 \pm 0.4$~\cite{Hiramatsu:2013qaa} and $A\simeq N$~\cite{Higaki:2016jjh,Hiramatsu:2012sc,Kawasaki:2014sqa} from simulations. 
The peak GW amplitude today is diluted by the cosmic expansion as
\begin{equation}
    \begin{aligned}
	& \!\!\!\!
    h^2\Omega_{\rm{GW}}^{\rm peak}\left(t_{0}\right)
    \\
    = & 6.45 \times 10^{-6} \varepsilon \left(\frac{\tilde{\epsilon}_{\rm{gw}}}{0.7}\right )\left( \frac{A}{10} \right )^2
    \left(\frac{g_{* s}\left(T_{\rm{ann}}\right)}{10}\right)^{-4/3} \\
    & \times \left(\frac{f}{100~\rm{TeV}}\right)^{6}\left(\frac{T_{\rm{ann}}}{0.1~\rm{GeV}}\right)^{-4}.
\end{aligned}
\end{equation}
We see that the DW annihilation in the clockwork framework can generate a GW amplitude of $\sim 10^{-6}$, which falls in the sensitivity of current PTA experiments.
Below the peak frequency, causality demands that the GW spectrum scale as $\propto (\nu/\nu_{\rm peak})^3$~\cite{Caprini:2009fx}. Above the peak frequency, on the other hand, numerical simulations~\cite{Hiramatsu:2013qaa} indicate a $\propto (\nu/\nu_{\rm peak})^{-1}$ scaling behavior instead.

{\it PTA data analysis and predictions.---} 
We carry out the standard Bayesian statistical analysis for the IPTA-DR2 dataset~\cite{IPTA:datalink} and the recent NG15 dataset~\cite{NANOGrav:2023gor} (see the Supplemental Material for more details). 
The Bayes estimators for the three input parameters $f/100~\rm TeV$, $\varepsilon$, and $N$ are $1.81\pm 0.21$, $0.50\pm 0.26$, and $12.21\pm 4.35$ ($1.76\pm 0.22$, $0.48\pm 0.26$, and $12.28\pm 4.35$) by fitting to the NG15 (IPTA-DR2) dataset. These parameter values are very natural in the clockwork axion framework and are also in good agreement with the prediction in our previous work~\cite{Chiang:2020aui} for the NG12 data. In addition to considering merely the DWs, in the Supplemental Material we also take into account potential astrophysical sources by including a power-law spectrum in our fit. We find that our interpretation is still supported by the NG15 data.

\begin{figure}
    \centering
    \includegraphics[width=90mm,angle=0]{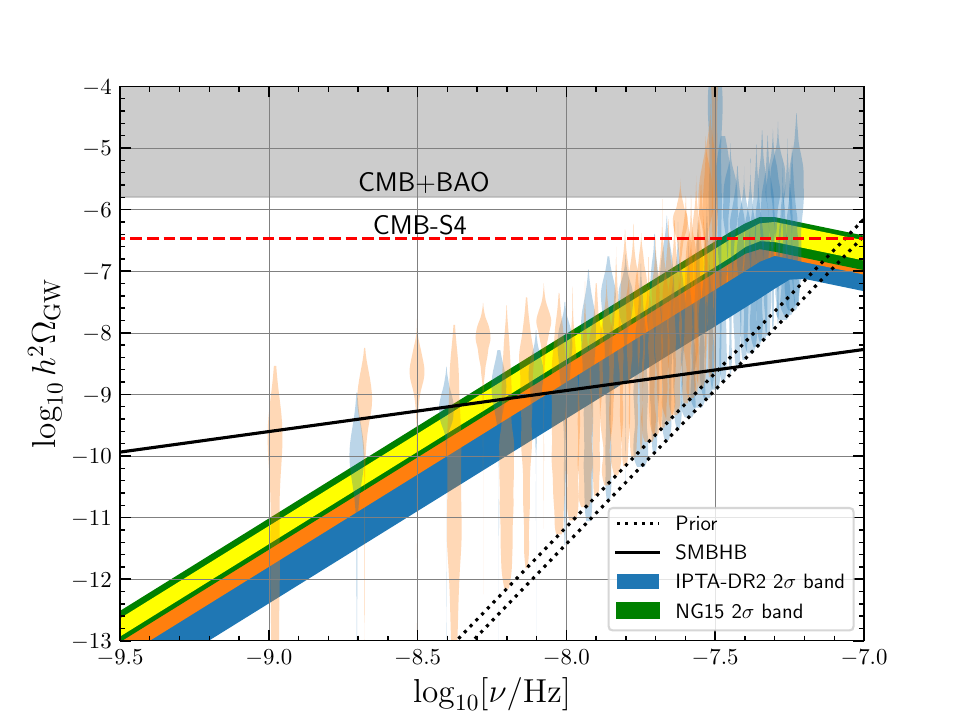}
    \caption{The black line denotes the SMBHB with the power-law spectrum $(A_{\rm BHB},~\gamma_{\rm BHB})=(-14.7,~13/3)$.
    The posteriors of the free spectrum for IPTA-DR2~\cite{Antoniadis:2022pcn} and NG15~\cite{NANOGrav:2023gor} are reproduced by the light-orange and light-blue violins, with the prior choices of lower limits shown by the dotted lines. 
    The yellow (green) region and orange (blue) region represent the 1~$\sigma$ (2~$\sigma$) uncertainty band by fitting to the NG15 and IPTA-DR2 datasets, respectively.}
    \label{fig:NANOGrav}
\end{figure} 

Note that although only the first 14 frequencies ($10^{-9.3}\lesssim \nu/{\rm Hz}\lesssim 10^{-7.6}$) of the NG15 dataset are adopted in the fit to avoid the possible high frequencies pulsar-intrinsic excess noise, we predict that the amplitude will continue to grow and have
a peak $h^{2}\Omega_{\rm GW}\simeq 10^{-6.6}-10^{-6.1}$ at $\nu\simeq 10^{-7.3}$~Hz (see Fig.~\ref{fig:NANOGrav} and Fig.~10 of~\cite{Chiang:2020aui}). 
This is because the peak frequency is entirely determined by the QCD instanton effect.
GWs can contribute to the radiation energy density and affect the expansion of the Universe.
This gives a constraint $h^2\Omega_{\mathrm{GW}}(t_0)\lesssim 5.6 \times 10^{-6} \Delta N_{\mathrm{eff}}$~\cite{Caprini:2018mtu}.
The current upper bound on the number of extra neutrino species at 95\% confidence level (CL) from the Planck observation is $\Delta N_{\rm eff}\leq 0.29$~\cite{Planck:2018vyg} (the grey region in Fig.~\ref{fig:NANOGrav}), which can be further tightened to 0.11 and 0.06 by
the upcoming Simons Observatory~\cite{SimonsObservatory:2018koc} and CMB-S4 experiment~\cite{CMB-S4:2022ght}.
We find that the predicted peak amplitude can generate detectable effects in the CMB-S4 experiment.
We also observe in Fig.~\ref{fig:prediction} that the predicted GW signals are detectable for future GW experiments in a wide frequency range.


\begin{figure}
    \centering
    \includegraphics[width=90mm,angle=0]{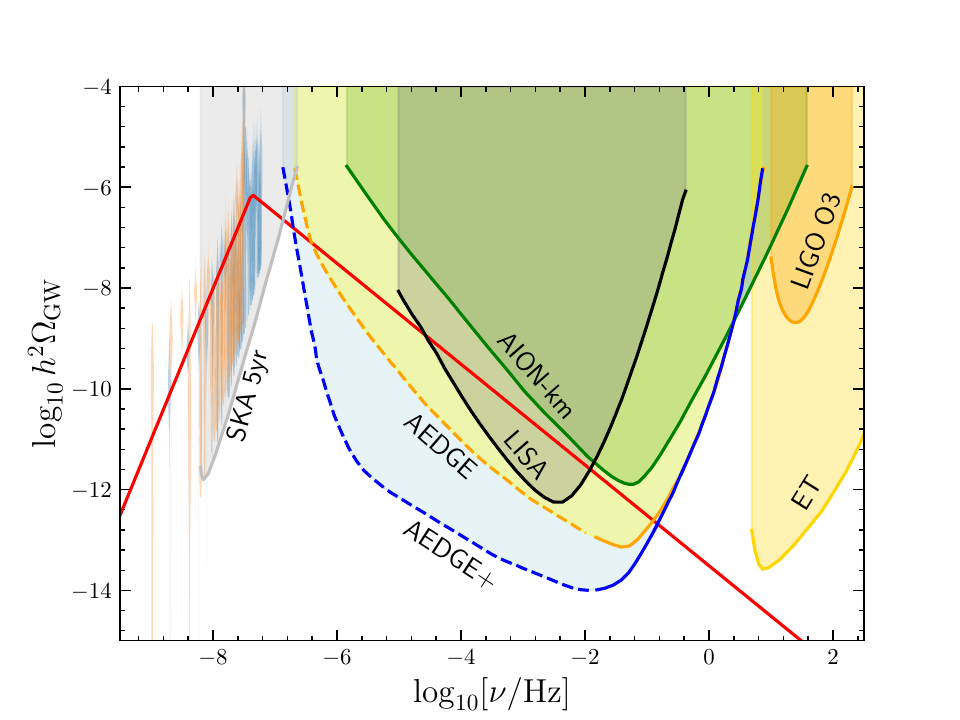}
    \caption{The SGWB sensitivity curves of SKA 5yr (silver)~\cite{Janssen:2014dka}, AION-km (light-green)~\cite{Badurina:2019hst}, AEDGE (yellow)~\cite{AEDGE:2019nxb}, AEDGE+ (light-blue), LISA (grey)~\cite{Caprini:2019egz}, LIGO O3 (orange)~\cite{Shoemaker:2019bqt}, Einstein telescope (ET, gold)~\cite{Maggiore:2019uih}. The red line corresponds to a GW signal with $(f,~\varepsilon,~N)=(180{\rm TeV},~0.5,~12)$. Note that the AEDGE sensitivities presented in~\cite{Badurina:2021rgt} do not consider many possible sources of instrumental noise.}
    \label{fig:prediction}
\end{figure} 

{\it Constraints.---}
We summarize various constraints in Fig.~\ref{fig:constraints}. 
Axions can be produced via the nucleon-nucleon axion bremsstrahlung in the core of a supernova and 
accelerate the stellar cooling.
The observation of neutrinos from SN~1987A excludes the light-blue region with $f_a\lesssim 10^9$~GeV~\cite{Mayle:1987as,Raffelt:1987yt,Turner:1987by}.
The emission of axions from the string-wall network is dominated by the late time decay at $t\sim t_{\rm QCD}$~\cite{Hiramatsu:2012gg}. The axion energy density from the decay of the scaling string is estimated as 
$\rho_{a,{\rm str}} \sim \pi f^2 H^2(t_{\rm QCD}) \ln (t_{\mathrm{QCD}} / t_{s})$, where $t_s$ is the string formation time.
The QCD axion obtains a mass $m_a$ and gets mixed with the massive axions $A_i$ with a mixing angle $\vartheta_i \sim q^Nm_a^2/m_{A}^2$ at the QCD scale.  
Then the QCD axion energy density from the collapse of the DW network is found to be 
$\rho_{a,{\rm wall}}\sim \sum_i \vartheta^2_i\rho_{w}(t_{\rm QCD})\sim 8Nm_a^4f_a^2H(t_{\rm QCD})/m_{A}^3$. We find that
the contributions from the scaling strings and collapsing walls to the cold axion relic abundance (with energy 
$\omega(t_{\rm QCD}) \sim H(t_{\rm QCD})$) and $\Delta N_{\rm eff}$ are both negligibly small.  
The most significant contribution to $\rho_a$ from the topological defects is found to be the oscillations of DW with a frequency 
$\omega(t_{\rm QCD}) \gg H(t_{\rm QCD})$ at the QCD scale, which optimistically gives 
$\rho_{a,{\rm osc}}\sim 0.1v_w^2\rho_{w}(t_{\rm QCD})$~\cite{Long:2018nsl}.
This radiation population contributes to cosmic expansion by $\Delta N_{\rm eff}\simeq 6.4\times 10^{-3}v_w^2\varepsilon^{1/2}\left( f/100~{\rm TeV} \right)^3$, where we adopt the wall velocity $v_w\sim 1$ for a conservative estimate.
The bound from the current Planck observation falls much behind the constraint from DW domination (light grey region).
The constraints from the upcoming Simons Observatory and CMB-S4 experiment are represented by the red dot-dashed and red dashed lines.

\begin{figure}
    \centering
    \includegraphics[width=90mm,angle=0]{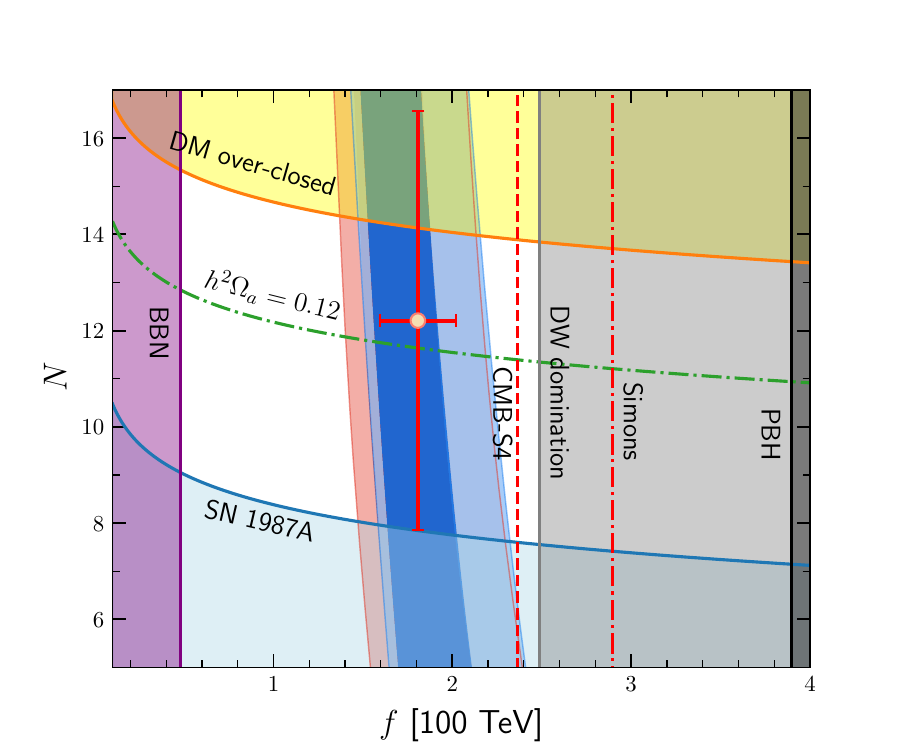}
    \caption{The blue and red vertical bands with $f\simeq 200~\rm TeV$ represent the 1 and 2$\sigma$ parameter bands 
    favored by NG15~\cite{NANOGrav:2023gor} and IPTA-DR2~\cite{Antoniadis:2022pcn} datasets, respectively. 
    The yellow region with $f_a\gtrsim 10^{12}$~GeV is excluded since the Universe is over-closed by the axion DM. }
    \label{fig:constraints}
\end{figure}

Since most of the energy from DW annihilation is poured into the SM plasma via the prompt decays of massive axion to the SM particles, we therefore require the annihilation to take place before the Big Bang Nucleosynthesis (BBN), i.e., $T_{\rm ann}>T_{\rm BBN}\simeq 5$~MeV, so that the successful BBN processes are not altered by the DW collapse (purple region). 
At the QCD scale, the closed DWs may collapse into massive PBHs~\cite{Ferrer:2018uiu}, whose lifetime can be longer than the Universe.
We find in the Supplemental Material that if $M_{\rm pl}^2\gtrsim 5\pi \sigma^2/V_{\rm bias}$, the PBH could be formed after the DW annihilation (black region). We also confirm that a negligible fraction $f_{\rm PBH}\lesssim 10^{-4}$ of DM consists of PBH formed in the collapse of DWs.  Finally, we note that the constraints from searches for axions in the laboratory are currently not competitive with the astrophysical bounds (see~\cite{DiLuzio:2020wdo} for a review).

{\it Phenomenology.---}
Axions start oscillations at $H(t)\sim m_a$ and generate a DM relic abundance
$\Omega_{a} h^2 \simeq 0.2\left(f_a/10^{11}~\mathrm{GeV}\right)^{7 /6} \left\langle\theta_a^2\right\rangle/(3\pi^2)$~\cite{Preskill:1982cy,Abbott:1982af,Dine:1982ah,Fox:2004kb,Choi:2020rgn},
where $\theta_a\equiv a/f_a$. 
Since the decay constant $f_a\simeq q^{N}f$ (with $q=3$) in the clockwork framework,
the large number $N=12.21\pm 4.35$ (point with error bars in Fig.~\ref{fig:constraints}) from the Bayes estimator naturally improves the axion decay constant to generate the correct axion DM relic abundance (dot-dashed green curve in Fig.~\ref{fig:constraints}).
Note that we have not taken into account the DM relic abundance observation in our fit. 
This is the first report on the novel correlation between the nano-Hz GWB and the DM relic abundance.
Such a correlation can be estimated by having $f_a\sim f^2/\Lambda_{\rm QCD}\sim 10^{11}$~GeV in the clockwork axion framework.
Equivalently, the DM relic abundance requires $f\simeq 200$~TeV, leading to the DW decays just before it would dominate the Universe.
Future telescope observations on stellar cooling~\cite{Ayala:2014pea,Giannotti:2017hny} and axion experiments may reveal the existence of axion~\cite{DiLuzio:2020wdo}.

The freeze-out of the massive particles can lead to changes in the relativistic DOF, and therefore, in the expansion rate $H(t)$ of the Universe. 
A rise in $H(t)$ after the annihilation of massive particles will dilute the subhorizon modes of the primordial GW spectrum, 
while the superhorizon modes are frozen and remain unaffected~\cite{Seto:2003kc,Watanabe:2006qe,Boyle:2005se,Chung:2010cb,Saikawa:2018rcs,Jinno:2011sw,Caldwell:2018giq}. 
Thus, the freeze-out of $N$ massive axions will leave an imprint on a primordial GW spectrum from inflation~\cite{Rubakov:1982df,Guzzetti:2016mkm,Caldwell:2017chz}, reheating/preheating~\cite{Khlebnikov:1997di,Easther:2006vd,Garcia-Bellido:2007fiu,Nakayama:2008wy}, or cosmic strings~\cite{Ellis:2020ena,Blasi:2020mfx,King:2020hyd}, and are detectable for future space-based interferometers, like LISA~\cite{Caprini:2019egz}, Taiji~\cite{Hu:2017mde}, and TianQin~\cite{TianQin:2015yph}.
The spectrum of superhorizon modes can be affected by the equation of state (EOS) when these modes enter the Hubble horizon 
during the QCD crossover, and therefore, leads to a slight departure from the $\nu^3$ behavior in the causality tail of the spectrum~\cite{Barenboim:2016mjm,Cai:2019cdl,Hook:2020phx,Franciolini:2023wjm}. 
Eq.~\eqref{eq:vpeak} indicates that the superhorizon mode begins around $10^{-8}$~Hz. Therefore PTA observations can be used to test the EOS at the QCD scale if the DW annihilation in our scenario is indeed the GW source for the PTAs.
Furthermore, the spontaneous breaking of the approximate $U(1)$ symmetries at $200$~TeV may lead to a cosmological first-order phase transition~\cite{Chiang:2020aui}, whose accompanying GW emissions with a characteristic spectrum peaked at even higher frequencies can be tested by the LIGO O3 run~\cite{Shoemaker:2019bqt} and ET~\cite{Maggiore:2019uih}.

{\it Conclusions and outlook.---}
For the first time, by fitting to the NG15 and IPTA-DR2 datasets, we found that
our predictions of the nano-Hz GW signal in Ref.~\cite{Chiang:2020aui} with a set of very natural parameters in the clockwork axion framework have been successfully and precisely confirmed by the recent PTA observations. Such success has not been observed in other works for the PTA observations, and thus strongly supports the novel mechanism of the QCD instanton-induced DW annihilation.
Furthermore, we found that the PTA data analysis naturally results in a correct DM relic abundance, which therefore indicates a strong correlation between the nano-Hz GWB and DM relic abundance. Such a correlation stems from the hierarchy between the bias and the symmetric parts of the potential. 
We for the first time proposed a novel relation $f_a\sim f^2/\Lambda_{\rm QCD}\sim 10^{11}$~GeV in the clockwork axion framework to estimate the correlation.
We showed that the GW signal has a peak of $h^{2}\Omega_{\rm GW}\simeq 10^{-6.6}-10^{-6.1}$ at about $50$~nHz and can be tested by PTAs and CMB-S4 experiment.
Moreover, we expected a slight departure from the $\nu^3$ scaling in the PTA spectrum at frequencies below about $ 10^{-8}$~Hz.
Our analysis may hint at the violation of Lorentz invariance in the extra dimension at $200$~TeV. Future GW observations will shed more light on model construction.

{\it Acknowledgments---}
We acknowledge the publicly available packages \texttt{enterprise}~\cite{Ellis:2020zenodo}, 
\texttt{enterprise\_extensions}~\cite{extensions2021},
\texttt{ceffyl}~\cite{Lamb:2023jls}, and \texttt{PTArcade}~\cite{Mitridate:2023oar}, 
where we have modeled the DW signal and implemented the Monte Carlo (MC) sampling with \texttt{PTMCMC}~\cite{Ellis:2017}.
BQL is supported in part by Zhejiang Provincial Natural Science Foundation of China under Grant No. LQ23A050002.
CWC is supported in part by the National Science and Technology Council under Grant No. NSTC-111-2112-M-002-018-MY3.
TL is supported in part by the National Key Research and Development Program of China Grant No. 2020YFC2201504, by the Projects No. 11875062, No. 11947302, No. 12047503, and No. 12275333 supported by the National Natural Science Foundation of China, by the Key Research Program of the Chinese Academy of Sciences, Grant NO. XDPB15, by the Scientific Instrument Developing Project of the Chinese Academy of Sciences, Grant No. YJKYYQ20190049, and by the International Partnership Program of Chinese Academy of Sciences for Grand Challenges, Grant No. 112311KYSB20210012.

\bibliography{reference}

\begin{thebibliography}{148}
\expandafter\ifx\csname natexlab\endcsname\relax\def\natexlab#1{#1}\fi
\expandafter\ifx\csname bibnamefont\endcsname\relax
  \def\bibnamefont#1{#1}\fi
\expandafter\ifx\csname bibfnamefont\endcsname\relax
  \def\bibfnamefont#1{#1}\fi
\expandafter\ifx\csname citenamefont\endcsname\relax
  \def\citenamefont#1{#1}\fi
\expandafter\ifx\csname url\endcsname\relax
  \def\url#1{\texttt{#1}}\fi
\expandafter\ifx\csname urlprefix\endcsname\relax\def\urlprefix{URL }\fi
\providecommand{\bibinfo}[2]{#2}
\providecommand{\eprint}[2][]{\url{#2}}

\bibitem[{\citenamefont{Zeldovich et~al.}(1974)\citenamefont{Zeldovich,
  Kobzarev, and Okun}}]{Zeldovich:1974uw}
\bibinfo{author}{\bibfnamefont{Y.~B.} \bibnamefont{Zeldovich}},
  \bibinfo{author}{\bibfnamefont{I.~Y.} \bibnamefont{Kobzarev}},
  \bibnamefont{and} \bibinfo{author}{\bibfnamefont{L.~B.} \bibnamefont{Okun}},
  \bibinfo{journal}{Zh. Eksp. Teor. Fiz.} \textbf{\bibinfo{volume}{67}},
  \bibinfo{pages}{3} (\bibinfo{year}{1974}).

\bibitem[{\citenamefont{Sikivie}(1982)}]{Sikivie:1982qv}
\bibinfo{author}{\bibfnamefont{P.}~\bibnamefont{Sikivie}},
  \bibinfo{journal}{Phys. Rev. Lett.} \textbf{\bibinfo{volume}{48}},
  \bibinfo{pages}{1156} (\bibinfo{year}{1982}).

\bibitem[{\citenamefont{Preskill et~al.}(1991)\citenamefont{Preskill, Trivedi,
  Wilczek, and Wise}}]{Preskill:1991kd}
\bibinfo{author}{\bibfnamefont{J.}~\bibnamefont{Preskill}},
  \bibinfo{author}{\bibfnamefont{S.~P.} \bibnamefont{Trivedi}},
  \bibinfo{author}{\bibfnamefont{F.}~\bibnamefont{Wilczek}}, \bibnamefont{and}
  \bibinfo{author}{\bibfnamefont{M.~B.} \bibnamefont{Wise}},
  \bibinfo{journal}{Nucl. Phys. B} \textbf{\bibinfo{volume}{363}},
  \bibinfo{pages}{207} (\bibinfo{year}{1991}).

\bibitem[{\citenamefont{Dine and Nelson}(1993)}]{Dine:1993yw}
\bibinfo{author}{\bibfnamefont{M.}~\bibnamefont{Dine}} \bibnamefont{and}
  \bibinfo{author}{\bibfnamefont{A.~E.} \bibnamefont{Nelson}},
  \bibinfo{journal}{Phys. Rev. D} \textbf{\bibinfo{volume}{48}},
  \bibinfo{pages}{1277} (\bibinfo{year}{1993}), \eprint{hep-ph/9303230}.

\bibitem[{\citenamefont{Dvali et~al.}(1995)\citenamefont{Dvali, Tavartkiladze,
  and Nanobashvili}}]{Dvali:1994wv}
\bibinfo{author}{\bibfnamefont{G.~R.} \bibnamefont{Dvali}},
  \bibinfo{author}{\bibfnamefont{Z.}~\bibnamefont{Tavartkiladze}},
  \bibnamefont{and}
  \bibinfo{author}{\bibfnamefont{J.}~\bibnamefont{Nanobashvili}},
  \bibinfo{journal}{Phys. Lett. B} \textbf{\bibinfo{volume}{352}},
  \bibinfo{pages}{214} (\bibinfo{year}{1995}), \eprint{hep-ph/9411387}.

\bibitem[{\citenamefont{Abel et~al.}(1995)\citenamefont{Abel, Sarkar, and
  White}}]{Abel:1995wk}
\bibinfo{author}{\bibfnamefont{S.~A.} \bibnamefont{Abel}},
  \bibinfo{author}{\bibfnamefont{S.}~\bibnamefont{Sarkar}}, \bibnamefont{and}
  \bibinfo{author}{\bibfnamefont{P.~L.} \bibnamefont{White}},
  \bibinfo{journal}{Nucl. Phys. B} \textbf{\bibinfo{volume}{454}},
  \bibinfo{pages}{663} (\bibinfo{year}{1995}), \eprint{hep-ph/9506359}.

\bibitem[{\citenamefont{Lee}(1988)}]{Lee:1988ge}
\bibinfo{author}{\bibfnamefont{K.-M.} \bibnamefont{Lee}},
  \bibinfo{journal}{Phys. Rev. Lett.} \textbf{\bibinfo{volume}{61}},
  \bibinfo{pages}{263} (\bibinfo{year}{1988}).

\bibitem[{\citenamefont{Abbott and Wise}(1989)}]{Abbott:1989jw}
\bibinfo{author}{\bibfnamefont{L.~F.} \bibnamefont{Abbott}} \bibnamefont{and}
  \bibinfo{author}{\bibfnamefont{M.~B.} \bibnamefont{Wise}},
  \bibinfo{journal}{Nucl. Phys. B} \textbf{\bibinfo{volume}{325}},
  \bibinfo{pages}{687} (\bibinfo{year}{1989}).

\bibitem[{\citenamefont{Coleman and Lee}(1990)}]{Coleman:1989zu}
\bibinfo{author}{\bibfnamefont{S.~R.} \bibnamefont{Coleman}} \bibnamefont{and}
  \bibinfo{author}{\bibfnamefont{K.-M.} \bibnamefont{Lee}},
  \bibinfo{journal}{Nucl. Phys. B} \textbf{\bibinfo{volume}{329}},
  \bibinfo{pages}{387} (\bibinfo{year}{1990}).

\bibitem[{\citenamefont{Gross et~al.}(1981)\citenamefont{Gross, Pisarski, and
  Yaffe}}]{Gross:1980br}
\bibinfo{author}{\bibfnamefont{D.~J.} \bibnamefont{Gross}},
  \bibinfo{author}{\bibfnamefont{R.~D.} \bibnamefont{Pisarski}},
  \bibnamefont{and} \bibinfo{author}{\bibfnamefont{L.~G.} \bibnamefont{Yaffe}},
  \bibinfo{journal}{Rev. Mod. Phys.} \textbf{\bibinfo{volume}{53}},
  \bibinfo{pages}{43} (\bibinfo{year}{1981}).

\bibitem[{\citenamefont{Sch\"afer and Shuryak}(1998)}]{Schafer:1996wv}
\bibinfo{author}{\bibfnamefont{T.}~\bibnamefont{Sch\"afer}} \bibnamefont{and}
  \bibinfo{author}{\bibfnamefont{E.~V.} \bibnamefont{Shuryak}},
  \bibinfo{journal}{Rev. Mod. Phys.} \textbf{\bibinfo{volume}{70}},
  \bibinfo{pages}{323} (\bibinfo{year}{1998}), \eprint{hep-ph/9610451}.

\bibitem[{\citenamefont{Banks and Dixon}(1988)}]{Banks:1988yz}
\bibinfo{author}{\bibfnamefont{T.}~\bibnamefont{Banks}} \bibnamefont{and}
  \bibinfo{author}{\bibfnamefont{L.~J.} \bibnamefont{Dixon}},
  \bibinfo{journal}{Nucl. Phys. B} \textbf{\bibinfo{volume}{307}},
  \bibinfo{pages}{93} (\bibinfo{year}{1988}).

\bibitem[{\citenamefont{Rai and Senjanovic}(1994)}]{Rai:1992xw}
\bibinfo{author}{\bibfnamefont{B.}~\bibnamefont{Rai}} \bibnamefont{and}
  \bibinfo{author}{\bibfnamefont{G.}~\bibnamefont{Senjanovic}},
  \bibinfo{journal}{Phys. Rev. D} \textbf{\bibinfo{volume}{49}},
  \bibinfo{pages}{2729} (\bibinfo{year}{1994}), \eprint{hep-ph/9301240}.

\bibitem[{\citenamefont{Banks and Seiberg}(2011)}]{Banks:2010zn}
\bibinfo{author}{\bibfnamefont{T.}~\bibnamefont{Banks}} \bibnamefont{and}
  \bibinfo{author}{\bibfnamefont{N.}~\bibnamefont{Seiberg}},
  \bibinfo{journal}{Phys. Rev. D} \textbf{\bibinfo{volume}{83}},
  \bibinfo{pages}{084019} (\bibinfo{year}{2011}), \eprint{1011.5120}.

\bibitem[{\citenamefont{Harlow and Ooguri}(2021)}]{Harlow:2018tng}
\bibinfo{author}{\bibfnamefont{D.}~\bibnamefont{Harlow}} \bibnamefont{and}
  \bibinfo{author}{\bibfnamefont{H.}~\bibnamefont{Ooguri}},
  \bibinfo{journal}{Commun. Math. Phys.} \textbf{\bibinfo{volume}{383}},
  \bibinfo{pages}{1669} (\bibinfo{year}{2021}), \eprint{1810.05338}.

\bibitem[{\citenamefont{Witten}(2018)}]{Witten:2017hdv}
\bibinfo{author}{\bibfnamefont{E.}~\bibnamefont{Witten}},
  \bibinfo{journal}{Nature Phys.} \textbf{\bibinfo{volume}{14}},
  \bibinfo{pages}{116} (\bibinfo{year}{2018}), \eprint{1710.01791}.

\bibitem[{\citenamefont{Antinucci et~al.}(2023)\citenamefont{Antinucci, Galati,
  Rizi, and Serone}}]{Antinucci:2023uzq}
\bibinfo{author}{\bibfnamefont{A.}~\bibnamefont{Antinucci}},
  \bibinfo{author}{\bibfnamefont{G.}~\bibnamefont{Galati}},
  \bibinfo{author}{\bibfnamefont{G.}~\bibnamefont{Rizi}}, \bibnamefont{and}
  \bibinfo{author}{\bibfnamefont{M.}~\bibnamefont{Serone}}
  (\bibinfo{year}{2023}), \eprint{2305.08911}.

\bibitem[{\citenamefont{Lattanzi and Valle}(2007)}]{Lattanzi:2007ux}
\bibinfo{author}{\bibfnamefont{M.}~\bibnamefont{Lattanzi}} \bibnamefont{and}
  \bibinfo{author}{\bibfnamefont{J.~W.~F.} \bibnamefont{Valle}},
  \bibinfo{journal}{Phys. Rev. Lett.} \textbf{\bibinfo{volume}{99}},
  \bibinfo{pages}{121301} (\bibinfo{year}{2007}), \eprint{0705.2406}.

\bibitem[{\citenamefont{King et~al.}(2023)\citenamefont{King, Roshan, Wang,
  White, and Yamazaki}}]{King:2023ayw}
\bibinfo{author}{\bibfnamefont{S.~F.} \bibnamefont{King}},
  \bibinfo{author}{\bibfnamefont{R.}~\bibnamefont{Roshan}},
  \bibinfo{author}{\bibfnamefont{X.}~\bibnamefont{Wang}},
  \bibinfo{author}{\bibfnamefont{G.}~\bibnamefont{White}}, \bibnamefont{and}
  \bibinfo{author}{\bibfnamefont{M.}~\bibnamefont{Yamazaki}}
  (\bibinfo{year}{2023}), \eprint{2308.03724}.

\bibitem[{\citenamefont{Craig et~al.}(2021)\citenamefont{Craig, Garcia~Garcia,
  Koszegi, and McCune}}]{Craig:2020bnv}
\bibinfo{author}{\bibfnamefont{N.}~\bibnamefont{Craig}},
  \bibinfo{author}{\bibfnamefont{I.}~\bibnamefont{Garcia~Garcia}},
  \bibinfo{author}{\bibfnamefont{G.}~\bibnamefont{Koszegi}}, \bibnamefont{and}
  \bibinfo{author}{\bibfnamefont{A.}~\bibnamefont{McCune}},
  \bibinfo{journal}{JHEP} \textbf{\bibinfo{volume}{09}}, \bibinfo{pages}{130}
  (\bibinfo{year}{2021}), \eprint{2012.13416}.

\bibitem[{\citenamefont{Chiang and Lu}(2021)}]{Chiang:2020aui}
\bibinfo{author}{\bibfnamefont{C.-W.} \bibnamefont{Chiang}} \bibnamefont{and}
  \bibinfo{author}{\bibfnamefont{B.-Q.} \bibnamefont{Lu}},
  \bibinfo{journal}{JCAP} \textbf{\bibinfo{volume}{05}}, \bibinfo{pages}{049}
  (\bibinfo{year}{2021}), \eprint{2012.14071}.

\bibitem[{\citenamefont{Arzoumanian et~al.}(2020)}]{NANOGrav:2020bcs}
\bibinfo{author}{\bibfnamefont{Z.}~\bibnamefont{Arzoumanian}}
  \bibnamefont{et~al.} (\bibinfo{collaboration}{NANOGrav}),
  \bibinfo{journal}{Astrophys. J. Lett.} \textbf{\bibinfo{volume}{905}},
  \bibinfo{pages}{L34} (\bibinfo{year}{2020}), \eprint{2009.04496}.

\bibitem[{\citenamefont{Ellis and Lewicki}(2021)}]{Ellis:2020ena}
\bibinfo{author}{\bibfnamefont{J.}~\bibnamefont{Ellis}} \bibnamefont{and}
  \bibinfo{author}{\bibfnamefont{M.}~\bibnamefont{Lewicki}},
  \bibinfo{journal}{Phys. Rev. Lett.} \textbf{\bibinfo{volume}{126}},
  \bibinfo{pages}{041304} (\bibinfo{year}{2021}), \eprint{2009.06555}.

\bibitem[{\citenamefont{Blasi et~al.}(2021)\citenamefont{Blasi, Brdar, and
  Schmitz}}]{Blasi:2020mfx}
\bibinfo{author}{\bibfnamefont{S.}~\bibnamefont{Blasi}},
  \bibinfo{author}{\bibfnamefont{V.}~\bibnamefont{Brdar}}, \bibnamefont{and}
  \bibinfo{author}{\bibfnamefont{K.}~\bibnamefont{Schmitz}},
  \bibinfo{journal}{Phys. Rev. Lett.} \textbf{\bibinfo{volume}{126}},
  \bibinfo{pages}{041305} (\bibinfo{year}{2021}), \eprint{2009.06607}.

\bibitem[{\citenamefont{King et~al.}(2021)\citenamefont{King, Pascoli, Turner,
  and Zhou}}]{King:2020hyd}
\bibinfo{author}{\bibfnamefont{S.~F.} \bibnamefont{King}},
  \bibinfo{author}{\bibfnamefont{S.}~\bibnamefont{Pascoli}},
  \bibinfo{author}{\bibfnamefont{J.}~\bibnamefont{Turner}}, \bibnamefont{and}
  \bibinfo{author}{\bibfnamefont{Y.-L.} \bibnamefont{Zhou}},
  \bibinfo{journal}{Phys. Rev. Lett.} \textbf{\bibinfo{volume}{126}},
  \bibinfo{pages}{021802} (\bibinfo{year}{2021}), \eprint{2005.13549}.

\bibitem[{\citenamefont{De~Luca et~al.}(2021)\citenamefont{De~Luca,
  Franciolini, and Riotto}}]{DeLuca:2020agl}
\bibinfo{author}{\bibfnamefont{V.}~\bibnamefont{De~Luca}},
  \bibinfo{author}{\bibfnamefont{G.}~\bibnamefont{Franciolini}},
  \bibnamefont{and} \bibinfo{author}{\bibfnamefont{A.}~\bibnamefont{Riotto}},
  \bibinfo{journal}{Phys. Rev. Lett.} \textbf{\bibinfo{volume}{126}},
  \bibinfo{pages}{041303} (\bibinfo{year}{2021}), \eprint{2009.08268}.

\bibitem[{\citenamefont{Vaskonen and Veerm\"ae}(2021)}]{Vaskonen:2020lbd}
\bibinfo{author}{\bibfnamefont{V.}~\bibnamefont{Vaskonen}} \bibnamefont{and}
  \bibinfo{author}{\bibfnamefont{H.}~\bibnamefont{Veerm\"ae}},
  \bibinfo{journal}{Phys. Rev. Lett.} \textbf{\bibinfo{volume}{126}},
  \bibinfo{pages}{051303} (\bibinfo{year}{2021}), \eprint{2009.07832}.

\bibitem[{\citenamefont{Agazie et~al.}(2023)}]{NANOGrav:2023gor}
\bibinfo{author}{\bibfnamefont{G.}~\bibnamefont{Agazie}} \bibnamefont{et~al.}
  (\bibinfo{collaboration}{NANOGrav}), \bibinfo{journal}{Astrophys. J. Lett.}
  \textbf{\bibinfo{volume}{951}}, \bibinfo{pages}{L8} (\bibinfo{year}{2023}),
  \eprint{2306.16213}.

\bibitem[{\citenamefont{Antoniadis et~al.}(2022)}]{Antoniadis:2022pcn}
\bibinfo{author}{\bibfnamefont{J.}~\bibnamefont{Antoniadis}}
  \bibnamefont{et~al.}, \bibinfo{journal}{Mon. Not. Roy. Astron. Soc.}
  \textbf{\bibinfo{volume}{510}}, \bibinfo{pages}{4873} (\bibinfo{year}{2022}),
  \eprint{2201.03980}.

\bibitem[{\citenamefont{Afzal et~al.}(2023)}]{NANOGrav:2023hvm}
\bibinfo{author}{\bibfnamefont{A.}~\bibnamefont{Afzal}} \bibnamefont{et~al.}
  (\bibinfo{collaboration}{NANOGrav}), \bibinfo{journal}{Astrophys. J. Lett.}
  \textbf{\bibinfo{volume}{951}}, \bibinfo{pages}{L11} (\bibinfo{year}{2023}),
  \eprint{2306.16219}.

\bibitem[{\citenamefont{Antoniadis et~al.}(2023)}]{EPTA:2023fyk}
\bibinfo{author}{\bibfnamefont{J.}~\bibnamefont{Antoniadis}}
  \bibnamefont{et~al.} (\bibinfo{collaboration}{EPTA}) (\bibinfo{year}{2023}),
  \eprint{2306.16214}.

\bibitem[{\citenamefont{Reardon et~al.}(2023)}]{Reardon:2023gzh}
\bibinfo{author}{\bibfnamefont{D.~J.} \bibnamefont{Reardon}}
  \bibnamefont{et~al.}, \bibinfo{journal}{Astrophys. J. Lett.}
  \textbf{\bibinfo{volume}{951}}, \bibinfo{pages}{L6} (\bibinfo{year}{2023}),
  \eprint{2306.16215}.

\bibitem[{\citenamefont{Xu et~al.}(2023)}]{Xu:2023wog}
\bibinfo{author}{\bibfnamefont{H.}~\bibnamefont{Xu}} \bibnamefont{et~al.},
  \bibinfo{journal}{Res. Astron. Astrophys.} \textbf{\bibinfo{volume}{23}},
  \bibinfo{pages}{075024} (\bibinfo{year}{2023}), \eprint{2306.16216}.

\bibitem[{\citenamefont{Guo et~al.}(2023)\citenamefont{Guo, Khlopov, Liu, Wu,
  Wu, and Zhu}}]{Guo:2023hyp}
\bibinfo{author}{\bibfnamefont{S.-Y.} \bibnamefont{Guo}},
  \bibinfo{author}{\bibfnamefont{M.}~\bibnamefont{Khlopov}},
  \bibinfo{author}{\bibfnamefont{X.}~\bibnamefont{Liu}},
  \bibinfo{author}{\bibfnamefont{L.}~\bibnamefont{Wu}},
  \bibinfo{author}{\bibfnamefont{Y.}~\bibnamefont{Wu}}, \bibnamefont{and}
  \bibinfo{author}{\bibfnamefont{B.}~\bibnamefont{Zhu}} (\bibinfo{year}{2023}),
  \eprint{2306.17022}.

\bibitem[{\citenamefont{Blasi et~al.}(2023{\natexlab{a}})\citenamefont{Blasi,
  Mariotti, Rase, and Sevrin}}]{Blasi:2023sej}
\bibinfo{author}{\bibfnamefont{S.}~\bibnamefont{Blasi}},
  \bibinfo{author}{\bibfnamefont{A.}~\bibnamefont{Mariotti}},
  \bibinfo{author}{\bibfnamefont{A.}~\bibnamefont{Rase}}, \bibnamefont{and}
  \bibinfo{author}{\bibfnamefont{A.}~\bibnamefont{Sevrin}}
  (\bibinfo{year}{2023}{\natexlab{a}}), \eprint{2306.17830}.

\bibitem[{\citenamefont{Du et~al.}(2023)\citenamefont{Du, Huang, Wang, and
  Zhang}}]{Du:2023qvj}
\bibinfo{author}{\bibfnamefont{X.~K.} \bibnamefont{Du}},
  \bibinfo{author}{\bibfnamefont{M.~X.} \bibnamefont{Huang}},
  \bibinfo{author}{\bibfnamefont{F.}~\bibnamefont{Wang}}, \bibnamefont{and}
  \bibinfo{author}{\bibfnamefont{Y.~K.} \bibnamefont{Zhang}}
  (\bibinfo{year}{2023}), \eprint{2307.02938}.

\bibitem[{\citenamefont{Bai et~al.}(2023)\citenamefont{Bai, Chen, and
  Korwar}}]{Bai:2023cqj}
\bibinfo{author}{\bibfnamefont{Y.}~\bibnamefont{Bai}},
  \bibinfo{author}{\bibfnamefont{T.-K.} \bibnamefont{Chen}}, \bibnamefont{and}
  \bibinfo{author}{\bibfnamefont{M.}~\bibnamefont{Korwar}}
  (\bibinfo{year}{2023}), \eprint{2306.17160}.

\bibitem[{\citenamefont{Zhang et~al.}(2023)\citenamefont{Zhang, Cai, Su, Wang,
  Yu, and Zhang}}]{Zhang:2023nrs}
\bibinfo{author}{\bibfnamefont{Z.}~\bibnamefont{Zhang}},
  \bibinfo{author}{\bibfnamefont{C.}~\bibnamefont{Cai}},
  \bibinfo{author}{\bibfnamefont{Y.-H.} \bibnamefont{Su}},
  \bibinfo{author}{\bibfnamefont{S.}~\bibnamefont{Wang}},
  \bibinfo{author}{\bibfnamefont{Z.-H.} \bibnamefont{Yu}}, \bibnamefont{and}
  \bibinfo{author}{\bibfnamefont{H.-H.} \bibnamefont{Zhang}}
  (\bibinfo{year}{2023}), \eprint{2307.11495}.

\bibitem[{\citenamefont{Ellis et~al.}(2023{\natexlab{a}})\citenamefont{Ellis,
  Lewicki, Lin, and Vaskonen}}]{Ellis:2023tsl}
\bibinfo{author}{\bibfnamefont{J.}~\bibnamefont{Ellis}},
  \bibinfo{author}{\bibfnamefont{M.}~\bibnamefont{Lewicki}},
  \bibinfo{author}{\bibfnamefont{C.}~\bibnamefont{Lin}}, \bibnamefont{and}
  \bibinfo{author}{\bibfnamefont{V.}~\bibnamefont{Vaskonen}}
  (\bibinfo{year}{2023}{\natexlab{a}}), \eprint{2306.17147}.

\bibitem[{\citenamefont{Wang et~al.}(2023{\natexlab{a}})\citenamefont{Wang,
  Lei, Jiao, Feng, and Fan}}]{Wang:2023len}
\bibinfo{author}{\bibfnamefont{Z.}~\bibnamefont{Wang}},
  \bibinfo{author}{\bibfnamefont{L.}~\bibnamefont{Lei}},
  \bibinfo{author}{\bibfnamefont{H.}~\bibnamefont{Jiao}},
  \bibinfo{author}{\bibfnamefont{L.}~\bibnamefont{Feng}}, \bibnamefont{and}
  \bibinfo{author}{\bibfnamefont{Y.-Z.} \bibnamefont{Fan}}
  (\bibinfo{year}{2023}{\natexlab{a}}), \eprint{2306.17150}.

\bibitem[{\citenamefont{Lazarides et~al.}(2023)\citenamefont{Lazarides, Maji,
  and Shafi}}]{Lazarides:2023ksx}
\bibinfo{author}{\bibfnamefont{G.}~\bibnamefont{Lazarides}},
  \bibinfo{author}{\bibfnamefont{R.}~\bibnamefont{Maji}}, \bibnamefont{and}
  \bibinfo{author}{\bibfnamefont{Q.}~\bibnamefont{Shafi}}
  (\bibinfo{year}{2023}), \eprint{2306.17788}.

\bibitem[{\citenamefont{Vagnozzi}(2023)}]{Vagnozzi:2023lwo}
\bibinfo{author}{\bibfnamefont{S.}~\bibnamefont{Vagnozzi}},
  \bibinfo{journal}{JHEAp} \textbf{\bibinfo{volume}{39}}, \bibinfo{pages}{81}
  (\bibinfo{year}{2023}), \eprint{2306.16912}.

\bibitem[{\citenamefont{Cai et~al.}(2023)\citenamefont{Cai, He, Ma, Yan, and
  Yuan}}]{Cai:2023dls}
\bibinfo{author}{\bibfnamefont{Y.-F.} \bibnamefont{Cai}},
  \bibinfo{author}{\bibfnamefont{X.-C.} \bibnamefont{He}},
  \bibinfo{author}{\bibfnamefont{X.}~\bibnamefont{Ma}},
  \bibinfo{author}{\bibfnamefont{S.-F.} \bibnamefont{Yan}}, \bibnamefont{and}
  \bibinfo{author}{\bibfnamefont{G.-W.} \bibnamefont{Yuan}}
  (\bibinfo{year}{2023}), \eprint{2306.17822}.

\bibitem[{\citenamefont{Wang et~al.}(2023{\natexlab{b}})\citenamefont{Wang,
  Zhao, Li, and Zhu}}]{Wang:2023ost}
\bibinfo{author}{\bibfnamefont{S.}~\bibnamefont{Wang}},
  \bibinfo{author}{\bibfnamefont{Z.-C.} \bibnamefont{Zhao}},
  \bibinfo{author}{\bibfnamefont{J.-P.} \bibnamefont{Li}}, \bibnamefont{and}
  \bibinfo{author}{\bibfnamefont{Q.-H.} \bibnamefont{Zhu}}
  (\bibinfo{year}{2023}{\natexlab{b}}), \eprint{2307.00572}.

\bibitem[{\citenamefont{Liu et~al.}(2023)\citenamefont{Liu, Chen, and
  Huang}}]{Liu:2023ymk}
\bibinfo{author}{\bibfnamefont{L.}~\bibnamefont{Liu}},
  \bibinfo{author}{\bibfnamefont{Z.-C.} \bibnamefont{Chen}}, \bibnamefont{and}
  \bibinfo{author}{\bibfnamefont{Q.-G.} \bibnamefont{Huang}}
  (\bibinfo{year}{2023}), \eprint{2307.01102}.

\bibitem[{\citenamefont{Yi et~al.}(2023)\citenamefont{Yi, Gao, Gong, Wang, and
  Zhang}}]{Yi:2023mbm}
\bibinfo{author}{\bibfnamefont{Z.}~\bibnamefont{Yi}},
  \bibinfo{author}{\bibfnamefont{Q.}~\bibnamefont{Gao}},
  \bibinfo{author}{\bibfnamefont{Y.}~\bibnamefont{Gong}},
  \bibinfo{author}{\bibfnamefont{Y.}~\bibnamefont{Wang}}, \bibnamefont{and}
  \bibinfo{author}{\bibfnamefont{F.}~\bibnamefont{Zhang}}
  (\bibinfo{year}{2023}), \eprint{2307.02467}.

\bibitem[{\citenamefont{Inomata et~al.}(2023)\citenamefont{Inomata, Kohri, and
  Terada}}]{Inomata:2023zup}
\bibinfo{author}{\bibfnamefont{K.}~\bibnamefont{Inomata}},
  \bibinfo{author}{\bibfnamefont{K.}~\bibnamefont{Kohri}}, \bibnamefont{and}
  \bibinfo{author}{\bibfnamefont{T.}~\bibnamefont{Terada}}
  (\bibinfo{year}{2023}), \eprint{2306.17834}.

\bibitem[{\citenamefont{Franciolini
  et~al.}(2023{\natexlab{a}})\citenamefont{Franciolini, Iovino, Vaskonen, and
  Veermae}}]{Franciolini:2023pbf}
\bibinfo{author}{\bibfnamefont{G.}~\bibnamefont{Franciolini}},
  \bibinfo{author}{\bibfnamefont{A.}~\bibnamefont{Iovino},
  \bibfnamefont{Junior.}},
  \bibinfo{author}{\bibfnamefont{V.}~\bibnamefont{Vaskonen}}, \bibnamefont{and}
  \bibinfo{author}{\bibfnamefont{H.}~\bibnamefont{Veermae}}
  (\bibinfo{year}{2023}{\natexlab{a}}), \eprint{2306.17149}.

\bibitem[{\citenamefont{Bhaumik et~al.}(2023)\citenamefont{Bhaumik, Jain, and
  Lewicki}}]{Bhaumik:2023wmw}
\bibinfo{author}{\bibfnamefont{N.}~\bibnamefont{Bhaumik}},
  \bibinfo{author}{\bibfnamefont{R.~K.} \bibnamefont{Jain}}, \bibnamefont{and}
  \bibinfo{author}{\bibfnamefont{M.}~\bibnamefont{Lewicki}}
  (\bibinfo{year}{2023}), \eprint{2308.07912}.

\bibitem[{\citenamefont{Han et~al.}(2023)\citenamefont{Han, Xie, Yang, and
  Zhang}}]{Han:2023olf}
\bibinfo{author}{\bibfnamefont{C.}~\bibnamefont{Han}},
  \bibinfo{author}{\bibfnamefont{K.-P.} \bibnamefont{Xie}},
  \bibinfo{author}{\bibfnamefont{J.~M.} \bibnamefont{Yang}}, \bibnamefont{and}
  \bibinfo{author}{\bibfnamefont{M.}~\bibnamefont{Zhang}}
  (\bibinfo{year}{2023}), \eprint{2306.16966}.

\bibitem[{\citenamefont{Megias et~al.}(2023)\citenamefont{Megias, Nardini, and
  Quiros}}]{Megias:2023kiy}
\bibinfo{author}{\bibfnamefont{E.}~\bibnamefont{Megias}},
  \bibinfo{author}{\bibfnamefont{G.}~\bibnamefont{Nardini}}, \bibnamefont{and}
  \bibinfo{author}{\bibfnamefont{M.}~\bibnamefont{Quiros}}
  (\bibinfo{year}{2023}), \eprint{2306.17071}.

\bibitem[{\citenamefont{Franciolini
  et~al.}(2023{\natexlab{b}})\citenamefont{Franciolini, Racco, and
  Rompineve}}]{Franciolini:2023wjm}
\bibinfo{author}{\bibfnamefont{G.}~\bibnamefont{Franciolini}},
  \bibinfo{author}{\bibfnamefont{D.}~\bibnamefont{Racco}}, \bibnamefont{and}
  \bibinfo{author}{\bibfnamefont{F.}~\bibnamefont{Rompineve}}
  (\bibinfo{year}{2023}{\natexlab{b}}), \eprint{2306.17136}.

\bibitem[{\citenamefont{Jiang et~al.}(2023)\citenamefont{Jiang, Yang, Ma, and
  Huang}}]{Jiang:2023qbm}
\bibinfo{author}{\bibfnamefont{S.}~\bibnamefont{Jiang}},
  \bibinfo{author}{\bibfnamefont{A.}~\bibnamefont{Yang}},
  \bibinfo{author}{\bibfnamefont{J.}~\bibnamefont{Ma}}, \bibnamefont{and}
  \bibinfo{author}{\bibfnamefont{F.~P.} \bibnamefont{Huang}}
  (\bibinfo{year}{2023}), \eprint{2306.17827}.

\bibitem[{\citenamefont{Zu et~al.}(2023)\citenamefont{Zu, Zhang, Li, Gu, Tsai,
  and Fan}}]{Zu:2023olm}
\bibinfo{author}{\bibfnamefont{L.}~\bibnamefont{Zu}},
  \bibinfo{author}{\bibfnamefont{C.}~\bibnamefont{Zhang}},
  \bibinfo{author}{\bibfnamefont{Y.-Y.} \bibnamefont{Li}},
  \bibinfo{author}{\bibfnamefont{Y.-C.} \bibnamefont{Gu}},
  \bibinfo{author}{\bibfnamefont{Y.-L.~S.} \bibnamefont{Tsai}},
  \bibnamefont{and} \bibinfo{author}{\bibfnamefont{Y.-Z.} \bibnamefont{Fan}}
  (\bibinfo{year}{2023}), \eprint{2306.16769}.

\bibitem[{\citenamefont{Xiao et~al.}(2023)\citenamefont{Xiao, Yang, and
  Zhang}}]{Xiao:2023dbb}
\bibinfo{author}{\bibfnamefont{Y.}~\bibnamefont{Xiao}},
  \bibinfo{author}{\bibfnamefont{J.~M.} \bibnamefont{Yang}}, \bibnamefont{and}
  \bibinfo{author}{\bibfnamefont{Y.}~\bibnamefont{Zhang}}
  (\bibinfo{year}{2023}), \eprint{2307.01072}.

\bibitem[{\citenamefont{Madge et~al.}(2023)\citenamefont{Madge, Morgante,
  Puchades-Ib\'a\~nez, Ramberg, Ratzinger, Schenk, and
  Schwaller}}]{Madge:2023cak}
\bibinfo{author}{\bibfnamefont{E.}~\bibnamefont{Madge}},
  \bibinfo{author}{\bibfnamefont{E.}~\bibnamefont{Morgante}},
  \bibinfo{author}{\bibfnamefont{C.}~\bibnamefont{Puchades-Ib\'a\~nez}},
  \bibinfo{author}{\bibfnamefont{N.}~\bibnamefont{Ramberg}},
  \bibinfo{author}{\bibfnamefont{W.}~\bibnamefont{Ratzinger}},
  \bibinfo{author}{\bibfnamefont{S.}~\bibnamefont{Schenk}}, \bibnamefont{and}
  \bibinfo{author}{\bibfnamefont{P.}~\bibnamefont{Schwaller}}
  (\bibinfo{year}{2023}), \eprint{2306.14856}.

\bibitem[{\citenamefont{Bian et~al.}(2023)\citenamefont{Bian, Ge, Shu, Wang,
  Yang, and Zong}}]{Bian:2023dnv}
\bibinfo{author}{\bibfnamefont{L.}~\bibnamefont{Bian}},
  \bibinfo{author}{\bibfnamefont{S.}~\bibnamefont{Ge}},
  \bibinfo{author}{\bibfnamefont{J.}~\bibnamefont{Shu}},
  \bibinfo{author}{\bibfnamefont{B.}~\bibnamefont{Wang}},
  \bibinfo{author}{\bibfnamefont{X.-Y.} \bibnamefont{Yang}}, \bibnamefont{and}
  \bibinfo{author}{\bibfnamefont{J.}~\bibnamefont{Zong}}
  (\bibinfo{year}{2023}), \eprint{2307.02376}.

\bibitem[{\citenamefont{Wu et~al.}(2023)\citenamefont{Wu, Chen, and
  Huang}}]{Wu:2023hsa}
\bibinfo{author}{\bibfnamefont{Y.-M.} \bibnamefont{Wu}},
  \bibinfo{author}{\bibfnamefont{Z.-C.} \bibnamefont{Chen}}, \bibnamefont{and}
  \bibinfo{author}{\bibfnamefont{Q.-G.} \bibnamefont{Huang}}
  (\bibinfo{year}{2023}), \eprint{2307.03141}.

\bibitem[{\citenamefont{Ellis et~al.}(2023{\natexlab{b}})\citenamefont{Ellis,
  Fairbairn, Franciolini, H\"utsi, Iovino, Lewicki, Raidal, Urrutia, Vaskonen,
  and Veerm\"ae}}]{Ellis:2023oxs}
\bibinfo{author}{\bibfnamefont{J.}~\bibnamefont{Ellis}},
  \bibinfo{author}{\bibfnamefont{M.}~\bibnamefont{Fairbairn}},
  \bibinfo{author}{\bibfnamefont{G.}~\bibnamefont{Franciolini}},
  \bibinfo{author}{\bibfnamefont{G.}~\bibnamefont{H\"utsi}},
  \bibinfo{author}{\bibfnamefont{A.}~\bibnamefont{Iovino}},
  \bibinfo{author}{\bibfnamefont{M.}~\bibnamefont{Lewicki}},
  \bibinfo{author}{\bibfnamefont{M.}~\bibnamefont{Raidal}},
  \bibinfo{author}{\bibfnamefont{J.}~\bibnamefont{Urrutia}},
  \bibinfo{author}{\bibfnamefont{V.}~\bibnamefont{Vaskonen}}, \bibnamefont{and}
  \bibinfo{author}{\bibfnamefont{H.}~\bibnamefont{Veerm\"ae}}
  (\bibinfo{year}{2023}{\natexlab{b}}), \eprint{2308.08546}.

\bibitem[{\citenamefont{Ade et~al.}(2016)}]{Planck:2015fie}
\bibinfo{author}{\bibfnamefont{P.~A.~R.} \bibnamefont{Ade}}
  \bibnamefont{et~al.} (\bibinfo{collaboration}{Planck}),
  \bibinfo{journal}{Astron. Astrophys.} \textbf{\bibinfo{volume}{594}},
  \bibinfo{pages}{A13} (\bibinfo{year}{2016}), \eprint{1502.01589}.

\bibitem[{\citenamefont{Choi et~al.}(2014)\citenamefont{Choi, Kim, and
  Yun}}]{Choi:2014rja}
\bibinfo{author}{\bibfnamefont{K.}~\bibnamefont{Choi}},
  \bibinfo{author}{\bibfnamefont{H.}~\bibnamefont{Kim}}, \bibnamefont{and}
  \bibinfo{author}{\bibfnamefont{S.}~\bibnamefont{Yun}},
  \bibinfo{journal}{Phys. Rev. D} \textbf{\bibinfo{volume}{90}},
  \bibinfo{pages}{023545} (\bibinfo{year}{2014}), \eprint{1404.6209}.

\bibitem[{\citenamefont{Choi and Im}(2016)}]{Choi:2015fiu}
\bibinfo{author}{\bibfnamefont{K.}~\bibnamefont{Choi}} \bibnamefont{and}
  \bibinfo{author}{\bibfnamefont{S.~H.} \bibnamefont{Im}},
  \bibinfo{journal}{JHEP} \textbf{\bibinfo{volume}{01}}, \bibinfo{pages}{149}
  (\bibinfo{year}{2016}), \eprint{1511.00132}.

\bibitem[{\citenamefont{Kaplan and Rattazzi}(2016)}]{Kaplan:2015fuy}
\bibinfo{author}{\bibfnamefont{D.~E.} \bibnamefont{Kaplan}} \bibnamefont{and}
  \bibinfo{author}{\bibfnamefont{R.}~\bibnamefont{Rattazzi}},
  \bibinfo{journal}{Phys. Rev. D} \textbf{\bibinfo{volume}{93}},
  \bibinfo{pages}{085007} (\bibinfo{year}{2016}), \eprint{1511.01827}.

\bibitem[{\citenamefont{Higaki et~al.}(2016{\natexlab{a}})\citenamefont{Higaki,
  Jeong, Kitajima, and Takahashi}}]{Higaki:2015jag}
\bibinfo{author}{\bibfnamefont{T.}~\bibnamefont{Higaki}},
  \bibinfo{author}{\bibfnamefont{K.~S.} \bibnamefont{Jeong}},
  \bibinfo{author}{\bibfnamefont{N.}~\bibnamefont{Kitajima}}, \bibnamefont{and}
  \bibinfo{author}{\bibfnamefont{F.}~\bibnamefont{Takahashi}},
  \bibinfo{journal}{Phys. Lett. B} \textbf{\bibinfo{volume}{755}},
  \bibinfo{pages}{13} (\bibinfo{year}{2016}{\natexlab{a}}),
  \eprint{1512.05295}.

\bibitem[{\citenamefont{Higaki et~al.}(2016{\natexlab{b}})\citenamefont{Higaki,
  Jeong, Kitajima, and Takahashi}}]{Higaki:2016yqk}
\bibinfo{author}{\bibfnamefont{T.}~\bibnamefont{Higaki}},
  \bibinfo{author}{\bibfnamefont{K.~S.} \bibnamefont{Jeong}},
  \bibinfo{author}{\bibfnamefont{N.}~\bibnamefont{Kitajima}}, \bibnamefont{and}
  \bibinfo{author}{\bibfnamefont{F.}~\bibnamefont{Takahashi}},
  \bibinfo{journal}{JHEP} \textbf{\bibinfo{volume}{06}}, \bibinfo{pages}{150}
  (\bibinfo{year}{2016}{\natexlab{b}}), \eprint{1603.02090}.

\bibitem[{\citenamefont{Higaki et~al.}(2016{\natexlab{c}})\citenamefont{Higaki,
  Jeong, Kitajima, Sekiguchi, and Takahashi}}]{Higaki:2016jjh}
\bibinfo{author}{\bibfnamefont{T.}~\bibnamefont{Higaki}},
  \bibinfo{author}{\bibfnamefont{K.~S.} \bibnamefont{Jeong}},
  \bibinfo{author}{\bibfnamefont{N.}~\bibnamefont{Kitajima}},
  \bibinfo{author}{\bibfnamefont{T.}~\bibnamefont{Sekiguchi}},
  \bibnamefont{and}
  \bibinfo{author}{\bibfnamefont{F.}~\bibnamefont{Takahashi}},
  \bibinfo{journal}{JHEP} \textbf{\bibinfo{volume}{08}}, \bibinfo{pages}{044}
  (\bibinfo{year}{2016}{\natexlab{c}}), \eprint{1606.05552}.

\bibitem[{\citenamefont{Giudice and McCullough}(2017)}]{Giudice:2016yja}
\bibinfo{author}{\bibfnamefont{G.~F.} \bibnamefont{Giudice}} \bibnamefont{and}
  \bibinfo{author}{\bibfnamefont{M.}~\bibnamefont{McCullough}},
  \bibinfo{journal}{JHEP} \textbf{\bibinfo{volume}{02}}, \bibinfo{pages}{036}
  (\bibinfo{year}{2017}), \eprint{1610.07962}.

\bibitem[{\citenamefont{Farina et~al.}(2017)\citenamefont{Farina, Pappadopulo,
  Rompineve, and Tesi}}]{Farina:2016tgd}
\bibinfo{author}{\bibfnamefont{M.}~\bibnamefont{Farina}},
  \bibinfo{author}{\bibfnamefont{D.}~\bibnamefont{Pappadopulo}},
  \bibinfo{author}{\bibfnamefont{F.}~\bibnamefont{Rompineve}},
  \bibnamefont{and} \bibinfo{author}{\bibfnamefont{A.}~\bibnamefont{Tesi}},
  \bibinfo{journal}{JHEP} \textbf{\bibinfo{volume}{01}}, \bibinfo{pages}{095}
  (\bibinfo{year}{2017}), \eprint{1611.09855}.

\bibitem[{\citenamefont{Coy et~al.}(2017)\citenamefont{Coy, Frigerio, and
  Ibe}}]{Coy:2017yex}
\bibinfo{author}{\bibfnamefont{R.}~\bibnamefont{Coy}},
  \bibinfo{author}{\bibfnamefont{M.}~\bibnamefont{Frigerio}}, \bibnamefont{and}
  \bibinfo{author}{\bibfnamefont{M.}~\bibnamefont{Ibe}},
  \bibinfo{journal}{JHEP} \textbf{\bibinfo{volume}{10}}, \bibinfo{pages}{002}
  (\bibinfo{year}{2017}), \eprint{1706.04529}.

\bibitem[{\citenamefont{Long}(2018)}]{Long:2018nsl}
\bibinfo{author}{\bibfnamefont{A.~J.} \bibnamefont{Long}},
  \bibinfo{journal}{JHEP} \textbf{\bibinfo{volume}{07}}, \bibinfo{pages}{066}
  (\bibinfo{year}{2018}), \eprint{1803.07086}.

\bibitem[{\citenamefont{Agrawal et~al.}(2018)\citenamefont{Agrawal, Fan, and
  Reece}}]{Agrawal:2018mkd}
\bibinfo{author}{\bibfnamefont{P.}~\bibnamefont{Agrawal}},
  \bibinfo{author}{\bibfnamefont{J.}~\bibnamefont{Fan}}, \bibnamefont{and}
  \bibinfo{author}{\bibfnamefont{M.}~\bibnamefont{Reece}},
  \bibinfo{journal}{JHEP} \textbf{\bibinfo{volume}{10}}, \bibinfo{pages}{193}
  (\bibinfo{year}{2018}), \eprint{1806.09621}.

\bibitem[{\citenamefont{Choi et~al.}(2018)\citenamefont{Choi, Im, and
  Shin}}]{Choi:2017ncj}
\bibinfo{author}{\bibfnamefont{K.}~\bibnamefont{Choi}},
  \bibinfo{author}{\bibfnamefont{S.~H.} \bibnamefont{Im}}, \bibnamefont{and}
  \bibinfo{author}{\bibfnamefont{C.~S.} \bibnamefont{Shin}},
  \bibinfo{journal}{JHEP} \textbf{\bibinfo{volume}{07}}, \bibinfo{pages}{113}
  (\bibinfo{year}{2018}), \eprint{1711.06228}.

\bibitem[{\citenamefont{Wood et~al.}(2023)\citenamefont{Wood, Saffin, and
  Avgoustidis}}]{Wood:2023lis}
\bibinfo{author}{\bibfnamefont{K.}~\bibnamefont{Wood}},
  \bibinfo{author}{\bibfnamefont{P.~M.} \bibnamefont{Saffin}},
  \bibnamefont{and}
  \bibinfo{author}{\bibfnamefont{A.}~\bibnamefont{Avgoustidis}},
  \bibinfo{journal}{JCAP} \textbf{\bibinfo{volume}{07}}, \bibinfo{pages}{062}
  (\bibinfo{year}{2023}), \eprint{2304.09205}.

\bibitem[{\citenamefont{Arkani-Hamed et~al.}(2001)\citenamefont{Arkani-Hamed,
  Cohen, and Georgi}}]{Arkani-Hamed:2001kyx}
\bibinfo{author}{\bibfnamefont{N.}~\bibnamefont{Arkani-Hamed}},
  \bibinfo{author}{\bibfnamefont{A.~G.} \bibnamefont{Cohen}}, \bibnamefont{and}
  \bibinfo{author}{\bibfnamefont{H.}~\bibnamefont{Georgi}},
  \bibinfo{journal}{Phys. Rev. Lett.} \textbf{\bibinfo{volume}{86}},
  \bibinfo{pages}{4757} (\bibinfo{year}{2001}), \eprint{hep-th/0104005}.

\bibitem[{\citenamefont{Arkani-Hamed et~al.}(2003)\citenamefont{Arkani-Hamed,
  Cheng, Creminelli, and Randall}}]{Arkani-Hamed:2003xts}
\bibinfo{author}{\bibfnamefont{N.}~\bibnamefont{Arkani-Hamed}},
  \bibinfo{author}{\bibfnamefont{H.-C.} \bibnamefont{Cheng}},
  \bibinfo{author}{\bibfnamefont{P.}~\bibnamefont{Creminelli}},
  \bibnamefont{and} \bibinfo{author}{\bibfnamefont{L.}~\bibnamefont{Randall}},
  \bibinfo{journal}{Phys. Rev. Lett.} \textbf{\bibinfo{volume}{90}},
  \bibinfo{pages}{221302} (\bibinfo{year}{2003}), \eprint{hep-th/0301218}.

\bibitem[{\citenamefont{Choi}(2004)}]{Choi:2003wr}
\bibinfo{author}{\bibfnamefont{K.}~\bibnamefont{Choi}}, \bibinfo{journal}{Phys.
  Rev. Lett.} \textbf{\bibinfo{volume}{92}}, \bibinfo{pages}{101602}
  (\bibinfo{year}{2004}), \eprint{hep-ph/0308024}.

\bibitem[{\citenamefont{Tanabashi et~al.}(2018)}]{ParticleDataGroup:2018ovx}
\bibinfo{author}{\bibfnamefont{M.}~\bibnamefont{Tanabashi}}
  \bibnamefont{et~al.} (\bibinfo{collaboration}{Particle Data Group}),
  \bibinfo{journal}{Phys. Rev. D} \textbf{\bibinfo{volume}{98}},
  \bibinfo{pages}{030001} (\bibinfo{year}{2018}).

\bibitem[{\citenamefont{Hiramatsu et~al.}(2014)\citenamefont{Hiramatsu,
  Kawasaki, and Saikawa}}]{Hiramatsu:2013qaa}
\bibinfo{author}{\bibfnamefont{T.}~\bibnamefont{Hiramatsu}},
  \bibinfo{author}{\bibfnamefont{M.}~\bibnamefont{Kawasaki}}, \bibnamefont{and}
  \bibinfo{author}{\bibfnamefont{K.}~\bibnamefont{Saikawa}},
  \bibinfo{journal}{JCAP} \textbf{\bibinfo{volume}{02}}, \bibinfo{pages}{031}
  (\bibinfo{year}{2014}), \eprint{1309.5001}.

\bibitem[{\citenamefont{Hiramatsu et~al.}(2013)\citenamefont{Hiramatsu,
  Kawasaki, Saikawa, and Sekiguchi}}]{Hiramatsu:2012sc}
\bibinfo{author}{\bibfnamefont{T.}~\bibnamefont{Hiramatsu}},
  \bibinfo{author}{\bibfnamefont{M.}~\bibnamefont{Kawasaki}},
  \bibinfo{author}{\bibfnamefont{K.}~\bibnamefont{Saikawa}}, \bibnamefont{and}
  \bibinfo{author}{\bibfnamefont{T.}~\bibnamefont{Sekiguchi}},
  \bibinfo{journal}{JCAP} \textbf{\bibinfo{volume}{01}}, \bibinfo{pages}{001}
  (\bibinfo{year}{2013}), \eprint{1207.3166}.

\bibitem[{\citenamefont{Kawasaki et~al.}(2015)\citenamefont{Kawasaki, Saikawa,
  and Sekiguchi}}]{Kawasaki:2014sqa}
\bibinfo{author}{\bibfnamefont{M.}~\bibnamefont{Kawasaki}},
  \bibinfo{author}{\bibfnamefont{K.}~\bibnamefont{Saikawa}}, \bibnamefont{and}
  \bibinfo{author}{\bibfnamefont{T.}~\bibnamefont{Sekiguchi}},
  \bibinfo{journal}{Phys. Rev. D} \textbf{\bibinfo{volume}{91}},
  \bibinfo{pages}{065014} (\bibinfo{year}{2015}), \eprint{1412.0789}.

\bibitem[{\citenamefont{Caprini et~al.}(2009)\citenamefont{Caprini, Durrer,
  Konstandin, and Servant}}]{Caprini:2009fx}
\bibinfo{author}{\bibfnamefont{C.}~\bibnamefont{Caprini}},
  \bibinfo{author}{\bibfnamefont{R.}~\bibnamefont{Durrer}},
  \bibinfo{author}{\bibfnamefont{T.}~\bibnamefont{Konstandin}},
  \bibnamefont{and} \bibinfo{author}{\bibfnamefont{G.}~\bibnamefont{Servant}},
  \bibinfo{journal}{Phys. Rev. D} \textbf{\bibinfo{volume}{79}},
  \bibinfo{pages}{083519} (\bibinfo{year}{2009}), \eprint{0901.1661}.

\bibitem[{\citenamefont{Ransom and the IPTADR2~team}()}]{IPTA:datalink}
\bibinfo{author}{\bibfnamefont{S.}~\bibnamefont{Ransom}} \bibnamefont{and}
  \bibinfo{author}{\bibnamefont{the IPTADR2~team}},
  \bibinfo{howpublished}{\url{https://gitlab.com/IPTA/DR2/-/tree/master/release/VersionB}}.

\bibitem[{\citenamefont{Caprini and Figueroa}(2018)}]{Caprini:2018mtu}
\bibinfo{author}{\bibfnamefont{C.}~\bibnamefont{Caprini}} \bibnamefont{and}
  \bibinfo{author}{\bibfnamefont{D.~G.} \bibnamefont{Figueroa}},
  \bibinfo{journal}{Class. Quant. Grav.} \textbf{\bibinfo{volume}{35}},
  \bibinfo{pages}{163001} (\bibinfo{year}{2018}), \eprint{1801.04268}.

\bibitem[{\citenamefont{Aghanim et~al.}(2020)}]{Planck:2018vyg}
\bibinfo{author}{\bibfnamefont{N.}~\bibnamefont{Aghanim}} \bibnamefont{et~al.}
  (\bibinfo{collaboration}{Planck}), \bibinfo{journal}{Astron. Astrophys.}
  \textbf{\bibinfo{volume}{641}}, \bibinfo{pages}{A6} (\bibinfo{year}{2020}),
  \bibinfo{note}{[Erratum: Astron.Astrophys. 652, C4 (2021)]},
  \eprint{1807.06209}.

\bibitem[{\citenamefont{Ade et~al.}(2019)}]{SimonsObservatory:2018koc}
\bibinfo{author}{\bibfnamefont{P.}~\bibnamefont{Ade}} \bibnamefont{et~al.}
  (\bibinfo{collaboration}{Simons Observatory}), \bibinfo{journal}{JCAP}
  \textbf{\bibinfo{volume}{02}}, \bibinfo{pages}{056} (\bibinfo{year}{2019}),
  \eprint{1808.07445}.

\bibitem[{\citenamefont{Abazajian et~al.}(2022)}]{CMB-S4:2022ght}
\bibinfo{author}{\bibfnamefont{K.}~\bibnamefont{Abazajian}}
  \bibnamefont{et~al.} (\bibinfo{collaboration}{CMB-S4})
  (\bibinfo{year}{2022}), \eprint{2203.08024}.

\bibitem[{\citenamefont{Janssen et~al.}(2015)}]{Janssen:2014dka}
\bibinfo{author}{\bibfnamefont{G.}~\bibnamefont{Janssen}} \bibnamefont{et~al.},
  \bibinfo{journal}{PoS} \textbf{\bibinfo{volume}{AASKA14}},
  \bibinfo{pages}{037} (\bibinfo{year}{2015}), \eprint{1501.00127}.

\bibitem[{\citenamefont{Badurina et~al.}(2020)}]{Badurina:2019hst}
\bibinfo{author}{\bibfnamefont{L.}~\bibnamefont{Badurina}}
  \bibnamefont{et~al.}, \bibinfo{journal}{JCAP} \textbf{\bibinfo{volume}{05}},
  \bibinfo{pages}{011} (\bibinfo{year}{2020}), \eprint{1911.11755}.

\bibitem[{\citenamefont{El-Neaj et~al.}(2020)}]{AEDGE:2019nxb}
\bibinfo{author}{\bibfnamefont{Y.~A.} \bibnamefont{El-Neaj}}
  \bibnamefont{et~al.} (\bibinfo{collaboration}{AEDGE}), \bibinfo{journal}{EPJ
  Quant. Technol.} \textbf{\bibinfo{volume}{7}}, \bibinfo{pages}{6}
  (\bibinfo{year}{2020}), \eprint{1908.00802}.

\bibitem[{\citenamefont{Caprini et~al.}(2020)}]{Caprini:2019egz}
\bibinfo{author}{\bibfnamefont{C.}~\bibnamefont{Caprini}} \bibnamefont{et~al.},
  \bibinfo{journal}{JCAP} \textbf{\bibinfo{volume}{03}}, \bibinfo{pages}{024}
  (\bibinfo{year}{2020}), \eprint{1910.13125}.

\bibitem[{\citenamefont{Shoemaker}(2019)}]{Shoemaker:2019bqt}
\bibinfo{author}{\bibfnamefont{D.}~\bibnamefont{Shoemaker}}
  (\bibinfo{collaboration}{LIGO Scientific}) (\bibinfo{year}{2019}),
  \eprint{1904.03187}.

\bibitem[{\citenamefont{Maggiore et~al.}(2020)}]{Maggiore:2019uih}
\bibinfo{author}{\bibfnamefont{M.}~\bibnamefont{Maggiore}}
  \bibnamefont{et~al.}, \bibinfo{journal}{JCAP} \textbf{\bibinfo{volume}{03}},
  \bibinfo{pages}{050} (\bibinfo{year}{2020}), \eprint{1912.02622}.

\bibitem[{\citenamefont{Badurina et~al.}(2021)\citenamefont{Badurina,
  Buchmueller, Ellis, Lewicki, McCabe, and Vaskonen}}]{Badurina:2021rgt}
\bibinfo{author}{\bibfnamefont{L.}~\bibnamefont{Badurina}},
  \bibinfo{author}{\bibfnamefont{O.}~\bibnamefont{Buchmueller}},
  \bibinfo{author}{\bibfnamefont{J.}~\bibnamefont{Ellis}},
  \bibinfo{author}{\bibfnamefont{M.}~\bibnamefont{Lewicki}},
  \bibinfo{author}{\bibfnamefont{C.}~\bibnamefont{McCabe}}, \bibnamefont{and}
  \bibinfo{author}{\bibfnamefont{V.}~\bibnamefont{Vaskonen}},
  \bibinfo{journal}{Phil. Trans. A. Math. Phys. Eng. Sci.}
  \textbf{\bibinfo{volume}{380}}, \bibinfo{pages}{20210060}
  (\bibinfo{year}{2021}), \eprint{2108.02468}.

\bibitem[{\citenamefont{Mayle et~al.}(1988)\citenamefont{Mayle, Wilson, Ellis,
  Olive, Schramm, and Steigman}}]{Mayle:1987as}
\bibinfo{author}{\bibfnamefont{R.}~\bibnamefont{Mayle}},
  \bibinfo{author}{\bibfnamefont{J.~R.} \bibnamefont{Wilson}},
  \bibinfo{author}{\bibfnamefont{J.~R.} \bibnamefont{Ellis}},
  \bibinfo{author}{\bibfnamefont{K.~A.} \bibnamefont{Olive}},
  \bibinfo{author}{\bibfnamefont{D.~N.} \bibnamefont{Schramm}},
  \bibnamefont{and} \bibinfo{author}{\bibfnamefont{G.}~\bibnamefont{Steigman}},
  \bibinfo{journal}{Phys. Lett. B} \textbf{\bibinfo{volume}{203}},
  \bibinfo{pages}{188} (\bibinfo{year}{1988}).

\bibitem[{\citenamefont{Raffelt and Seckel}(1988)}]{Raffelt:1987yt}
\bibinfo{author}{\bibfnamefont{G.}~\bibnamefont{Raffelt}} \bibnamefont{and}
  \bibinfo{author}{\bibfnamefont{D.}~\bibnamefont{Seckel}},
  \bibinfo{journal}{Phys. Rev. Lett.} \textbf{\bibinfo{volume}{60}},
  \bibinfo{pages}{1793} (\bibinfo{year}{1988}).

\bibitem[{\citenamefont{Turner}(1988)}]{Turner:1987by}
\bibinfo{author}{\bibfnamefont{M.~S.} \bibnamefont{Turner}},
  \bibinfo{journal}{Phys. Rev. Lett.} \textbf{\bibinfo{volume}{60}},
  \bibinfo{pages}{1797} (\bibinfo{year}{1988}).

\bibitem[{\citenamefont{Hiramatsu et~al.}(2012)\citenamefont{Hiramatsu,
  Kawasaki, Saikawa, and Sekiguchi}}]{Hiramatsu:2012gg}
\bibinfo{author}{\bibfnamefont{T.}~\bibnamefont{Hiramatsu}},
  \bibinfo{author}{\bibfnamefont{M.}~\bibnamefont{Kawasaki}},
  \bibinfo{author}{\bibfnamefont{K.}~\bibnamefont{Saikawa}}, \bibnamefont{and}
  \bibinfo{author}{\bibfnamefont{T.}~\bibnamefont{Sekiguchi}},
  \bibinfo{journal}{Phys. Rev. D} \textbf{\bibinfo{volume}{85}},
  \bibinfo{pages}{105020} (\bibinfo{year}{2012}), \bibinfo{note}{[Erratum:
  Phys.Rev.D 86, 089902 (2012)]}, \eprint{1202.5851}.

\bibitem[{\citenamefont{Ferrer et~al.}(2019)\citenamefont{Ferrer, Masso,
  Panico, Pujolas, and Rompineve}}]{Ferrer:2018uiu}
\bibinfo{author}{\bibfnamefont{F.}~\bibnamefont{Ferrer}},
  \bibinfo{author}{\bibfnamefont{E.}~\bibnamefont{Masso}},
  \bibinfo{author}{\bibfnamefont{G.}~\bibnamefont{Panico}},
  \bibinfo{author}{\bibfnamefont{O.}~\bibnamefont{Pujolas}}, \bibnamefont{and}
  \bibinfo{author}{\bibfnamefont{F.}~\bibnamefont{Rompineve}},
  \bibinfo{journal}{Phys. Rev. Lett.} \textbf{\bibinfo{volume}{122}},
  \bibinfo{pages}{101301} (\bibinfo{year}{2019}), \eprint{1807.01707}.

\bibitem[{\citenamefont{Di~Luzio et~al.}(2020)\citenamefont{Di~Luzio,
  Giannotti, Nardi, and Visinelli}}]{DiLuzio:2020wdo}
\bibinfo{author}{\bibfnamefont{L.}~\bibnamefont{Di~Luzio}},
  \bibinfo{author}{\bibfnamefont{M.}~\bibnamefont{Giannotti}},
  \bibinfo{author}{\bibfnamefont{E.}~\bibnamefont{Nardi}}, \bibnamefont{and}
  \bibinfo{author}{\bibfnamefont{L.}~\bibnamefont{Visinelli}},
  \bibinfo{journal}{Phys. Rept.} \textbf{\bibinfo{volume}{870}},
  \bibinfo{pages}{1} (\bibinfo{year}{2020}), \eprint{2003.01100}.

\bibitem[{\citenamefont{Preskill et~al.}(1983)\citenamefont{Preskill, Wise, and
  Wilczek}}]{Preskill:1982cy}
\bibinfo{author}{\bibfnamefont{J.}~\bibnamefont{Preskill}},
  \bibinfo{author}{\bibfnamefont{M.~B.} \bibnamefont{Wise}}, \bibnamefont{and}
  \bibinfo{author}{\bibfnamefont{F.}~\bibnamefont{Wilczek}},
  \bibinfo{journal}{Phys. Lett. B} \textbf{\bibinfo{volume}{120}},
  \bibinfo{pages}{127} (\bibinfo{year}{1983}).

\bibitem[{\citenamefont{Abbott and Sikivie}(1983)}]{Abbott:1982af}
\bibinfo{author}{\bibfnamefont{L.~F.} \bibnamefont{Abbott}} \bibnamefont{and}
  \bibinfo{author}{\bibfnamefont{P.}~\bibnamefont{Sikivie}},
  \bibinfo{journal}{Phys. Lett. B} \textbf{\bibinfo{volume}{120}},
  \bibinfo{pages}{133} (\bibinfo{year}{1983}).

\bibitem[{\citenamefont{Dine and Fischler}(1983)}]{Dine:1982ah}
\bibinfo{author}{\bibfnamefont{M.}~\bibnamefont{Dine}} \bibnamefont{and}
  \bibinfo{author}{\bibfnamefont{W.}~\bibnamefont{Fischler}},
  \bibinfo{journal}{Phys. Lett. B} \textbf{\bibinfo{volume}{120}},
  \bibinfo{pages}{137} (\bibinfo{year}{1983}).

\bibitem[{\citenamefont{Fox et~al.}(2004)\citenamefont{Fox, Pierce, and
  Thomas}}]{Fox:2004kb}
\bibinfo{author}{\bibfnamefont{P.}~\bibnamefont{Fox}},
  \bibinfo{author}{\bibfnamefont{A.}~\bibnamefont{Pierce}}, \bibnamefont{and}
  \bibinfo{author}{\bibfnamefont{S.~D.} \bibnamefont{Thomas}}
  (\bibinfo{year}{2004}), \eprint{hep-th/0409059}.

\bibitem[{\citenamefont{Choi et~al.}(2021)\citenamefont{Choi, Im, and
  Sub~Shin}}]{Choi:2020rgn}
\bibinfo{author}{\bibfnamefont{K.}~\bibnamefont{Choi}},
  \bibinfo{author}{\bibfnamefont{S.~H.} \bibnamefont{Im}}, \bibnamefont{and}
  \bibinfo{author}{\bibfnamefont{C.}~\bibnamefont{Sub~Shin}},
  \bibinfo{journal}{Ann. Rev. Nucl. Part. Sci.} \textbf{\bibinfo{volume}{71}},
  \bibinfo{pages}{225} (\bibinfo{year}{2021}), \eprint{2012.05029}.

\bibitem[{\citenamefont{Ayala et~al.}(2014)\citenamefont{Ayala, Dom\'\i{}nguez,
  Giannotti, Mirizzi, and Straniero}}]{Ayala:2014pea}
\bibinfo{author}{\bibfnamefont{A.}~\bibnamefont{Ayala}},
  \bibinfo{author}{\bibfnamefont{I.}~\bibnamefont{Dom\'\i{}nguez}},
  \bibinfo{author}{\bibfnamefont{M.}~\bibnamefont{Giannotti}},
  \bibinfo{author}{\bibfnamefont{A.}~\bibnamefont{Mirizzi}}, \bibnamefont{and}
  \bibinfo{author}{\bibfnamefont{O.}~\bibnamefont{Straniero}},
  \bibinfo{journal}{Phys. Rev. Lett.} \textbf{\bibinfo{volume}{113}},
  \bibinfo{pages}{191302} (\bibinfo{year}{2014}), \eprint{1406.6053}.

\bibitem[{\citenamefont{Giannotti et~al.}(2017)\citenamefont{Giannotti,
  Irastorza, Redondo, Ringwald, and Saikawa}}]{Giannotti:2017hny}
\bibinfo{author}{\bibfnamefont{M.}~\bibnamefont{Giannotti}},
  \bibinfo{author}{\bibfnamefont{I.~G.} \bibnamefont{Irastorza}},
  \bibinfo{author}{\bibfnamefont{J.}~\bibnamefont{Redondo}},
  \bibinfo{author}{\bibfnamefont{A.}~\bibnamefont{Ringwald}}, \bibnamefont{and}
  \bibinfo{author}{\bibfnamefont{K.}~\bibnamefont{Saikawa}},
  \bibinfo{journal}{JCAP} \textbf{\bibinfo{volume}{10}}, \bibinfo{pages}{010}
  (\bibinfo{year}{2017}), \eprint{1708.02111}.

\bibitem[{\citenamefont{Seto and Yokoyama}(2003)}]{Seto:2003kc}
\bibinfo{author}{\bibfnamefont{N.}~\bibnamefont{Seto}} \bibnamefont{and}
  \bibinfo{author}{\bibfnamefont{J.}~\bibnamefont{Yokoyama}},
  \bibinfo{journal}{J. Phys. Soc. Jap.} \textbf{\bibinfo{volume}{72}},
  \bibinfo{pages}{3082} (\bibinfo{year}{2003}), \eprint{gr-qc/0305096}.

\bibitem[{\citenamefont{Watanabe and Komatsu}(2006)}]{Watanabe:2006qe}
\bibinfo{author}{\bibfnamefont{Y.}~\bibnamefont{Watanabe}} \bibnamefont{and}
  \bibinfo{author}{\bibfnamefont{E.}~\bibnamefont{Komatsu}},
  \bibinfo{journal}{Phys. Rev. D} \textbf{\bibinfo{volume}{73}},
  \bibinfo{pages}{123515} (\bibinfo{year}{2006}), \eprint{astro-ph/0604176}.

\bibitem[{\citenamefont{Boyle and Steinhardt}(2008)}]{Boyle:2005se}
\bibinfo{author}{\bibfnamefont{L.~A.} \bibnamefont{Boyle}} \bibnamefont{and}
  \bibinfo{author}{\bibfnamefont{P.~J.} \bibnamefont{Steinhardt}},
  \bibinfo{journal}{Phys. Rev. D} \textbf{\bibinfo{volume}{77}},
  \bibinfo{pages}{063504} (\bibinfo{year}{2008}), \eprint{astro-ph/0512014}.

\bibitem[{\citenamefont{Chung and Zhou}(2010)}]{Chung:2010cb}
\bibinfo{author}{\bibfnamefont{D.~J.~H.} \bibnamefont{Chung}} \bibnamefont{and}
  \bibinfo{author}{\bibfnamefont{P.}~\bibnamefont{Zhou}},
  \bibinfo{journal}{Phys. Rev. D} \textbf{\bibinfo{volume}{82}},
  \bibinfo{pages}{024027} (\bibinfo{year}{2010}), \eprint{1003.2462}.

\bibitem[{\citenamefont{Saikawa and Shirai}(2018)}]{Saikawa:2018rcs}
\bibinfo{author}{\bibfnamefont{K.}~\bibnamefont{Saikawa}} \bibnamefont{and}
  \bibinfo{author}{\bibfnamefont{S.}~\bibnamefont{Shirai}},
  \bibinfo{journal}{JCAP} \textbf{\bibinfo{volume}{05}}, \bibinfo{pages}{035}
  (\bibinfo{year}{2018}), \eprint{1803.01038}.

\bibitem[{\citenamefont{Jinno et~al.}(2012)\citenamefont{Jinno, Moroi, and
  Nakayama}}]{Jinno:2011sw}
\bibinfo{author}{\bibfnamefont{R.}~\bibnamefont{Jinno}},
  \bibinfo{author}{\bibfnamefont{T.}~\bibnamefont{Moroi}}, \bibnamefont{and}
  \bibinfo{author}{\bibfnamefont{K.}~\bibnamefont{Nakayama}},
  \bibinfo{journal}{Phys. Lett. B} \textbf{\bibinfo{volume}{713}},
  \bibinfo{pages}{129} (\bibinfo{year}{2012}), \eprint{1112.0084}.

\bibitem[{\citenamefont{Caldwell et~al.}(2019)\citenamefont{Caldwell, Smith,
  and Walker}}]{Caldwell:2018giq}
\bibinfo{author}{\bibfnamefont{R.~R.} \bibnamefont{Caldwell}},
  \bibinfo{author}{\bibfnamefont{T.~L.} \bibnamefont{Smith}}, \bibnamefont{and}
  \bibinfo{author}{\bibfnamefont{D.~G.~E.} \bibnamefont{Walker}},
  \bibinfo{journal}{Phys. Rev. D} \textbf{\bibinfo{volume}{100}},
  \bibinfo{pages}{043513} (\bibinfo{year}{2019}), \eprint{1812.07577}.

\bibitem[{\citenamefont{Rubakov et~al.}(1982)\citenamefont{Rubakov, Sazhin, and
  Veryaskin}}]{Rubakov:1982df}
\bibinfo{author}{\bibfnamefont{V.~A.} \bibnamefont{Rubakov}},
  \bibinfo{author}{\bibfnamefont{M.~V.} \bibnamefont{Sazhin}},
  \bibnamefont{and} \bibinfo{author}{\bibfnamefont{A.~V.}
  \bibnamefont{Veryaskin}}, \bibinfo{journal}{Phys. Lett. B}
  \textbf{\bibinfo{volume}{115}}, \bibinfo{pages}{189} (\bibinfo{year}{1982}).

\bibitem[{\citenamefont{Guzzetti et~al.}(2016)\citenamefont{Guzzetti, Bartolo,
  Liguori, and Matarrese}}]{Guzzetti:2016mkm}
\bibinfo{author}{\bibfnamefont{M.~C.} \bibnamefont{Guzzetti}},
  \bibinfo{author}{\bibfnamefont{N.}~\bibnamefont{Bartolo}},
  \bibinfo{author}{\bibfnamefont{M.}~\bibnamefont{Liguori}}, \bibnamefont{and}
  \bibinfo{author}{\bibfnamefont{S.}~\bibnamefont{Matarrese}},
  \bibinfo{journal}{Riv. Nuovo Cim.} \textbf{\bibinfo{volume}{39}},
  \bibinfo{pages}{399} (\bibinfo{year}{2016}), \eprint{1605.01615}.

\bibitem[{\citenamefont{Caldwell and Devulder}(2018)}]{Caldwell:2017chz}
\bibinfo{author}{\bibfnamefont{R.~R.} \bibnamefont{Caldwell}} \bibnamefont{and}
  \bibinfo{author}{\bibfnamefont{C.}~\bibnamefont{Devulder}},
  \bibinfo{journal}{Phys. Rev. D} \textbf{\bibinfo{volume}{97}},
  \bibinfo{pages}{023532} (\bibinfo{year}{2018}), \eprint{1706.03765}.

\bibitem[{\citenamefont{Khlebnikov and Tkachev}(1997)}]{Khlebnikov:1997di}
\bibinfo{author}{\bibfnamefont{S.~Y.} \bibnamefont{Khlebnikov}}
  \bibnamefont{and} \bibinfo{author}{\bibfnamefont{I.~I.}
  \bibnamefont{Tkachev}}, \bibinfo{journal}{Phys. Rev. D}
  \textbf{\bibinfo{volume}{56}}, \bibinfo{pages}{653} (\bibinfo{year}{1997}),
  \eprint{hep-ph/9701423}.

\bibitem[{\citenamefont{Easther et~al.}(2007)\citenamefont{Easther, Giblin, and
  Lim}}]{Easther:2006vd}
\bibinfo{author}{\bibfnamefont{R.}~\bibnamefont{Easther}},
  \bibinfo{author}{\bibfnamefont{J.~T.} \bibnamefont{Giblin},
  \bibfnamefont{Jr.}}, \bibnamefont{and} \bibinfo{author}{\bibfnamefont{E.~A.}
  \bibnamefont{Lim}}, \bibinfo{journal}{Phys. Rev. Lett.}
  \textbf{\bibinfo{volume}{99}}, \bibinfo{pages}{221301}
  (\bibinfo{year}{2007}), \eprint{astro-ph/0612294}.

\bibitem[{\citenamefont{Garcia-Bellido
  et~al.}(2008)\citenamefont{Garcia-Bellido, Figueroa, and
  Sastre}}]{Garcia-Bellido:2007fiu}
\bibinfo{author}{\bibfnamefont{J.}~\bibnamefont{Garcia-Bellido}},
  \bibinfo{author}{\bibfnamefont{D.~G.} \bibnamefont{Figueroa}},
  \bibnamefont{and} \bibinfo{author}{\bibfnamefont{A.}~\bibnamefont{Sastre}},
  \bibinfo{journal}{Phys. Rev. D} \textbf{\bibinfo{volume}{77}},
  \bibinfo{pages}{043517} (\bibinfo{year}{2008}), \eprint{0707.0839}.

\bibitem[{\citenamefont{Nakayama et~al.}(2008)\citenamefont{Nakayama, Saito,
  Suwa, and Yokoyama}}]{Nakayama:2008wy}
\bibinfo{author}{\bibfnamefont{K.}~\bibnamefont{Nakayama}},
  \bibinfo{author}{\bibfnamefont{S.}~\bibnamefont{Saito}},
  \bibinfo{author}{\bibfnamefont{Y.}~\bibnamefont{Suwa}}, \bibnamefont{and}
  \bibinfo{author}{\bibfnamefont{J.}~\bibnamefont{Yokoyama}},
  \bibinfo{journal}{JCAP} \textbf{\bibinfo{volume}{06}}, \bibinfo{pages}{020}
  (\bibinfo{year}{2008}), \eprint{0804.1827}.

\bibitem[{\citenamefont{Hu and Wu}(2017)}]{Hu:2017mde}
\bibinfo{author}{\bibfnamefont{W.-R.} \bibnamefont{Hu}} \bibnamefont{and}
  \bibinfo{author}{\bibfnamefont{Y.-L.} \bibnamefont{Wu}},
  \bibinfo{journal}{Natl. Sci. Rev.} \textbf{\bibinfo{volume}{4}},
  \bibinfo{pages}{685} (\bibinfo{year}{2017}).

\bibitem[{\citenamefont{Luo et~al.}(2016)}]{TianQin:2015yph}
\bibinfo{author}{\bibfnamefont{J.}~\bibnamefont{Luo}} \bibnamefont{et~al.}
  (\bibinfo{collaboration}{TianQin}), \bibinfo{journal}{Class. Quant. Grav.}
  \textbf{\bibinfo{volume}{33}}, \bibinfo{pages}{035010}
  (\bibinfo{year}{2016}), \eprint{1512.02076}.

\bibitem[{\citenamefont{Barenboim and Park}(2016)}]{Barenboim:2016mjm}
\bibinfo{author}{\bibfnamefont{G.}~\bibnamefont{Barenboim}} \bibnamefont{and}
  \bibinfo{author}{\bibfnamefont{W.-I.} \bibnamefont{Park}},
  \bibinfo{journal}{Phys. Lett. B} \textbf{\bibinfo{volume}{759}},
  \bibinfo{pages}{430} (\bibinfo{year}{2016}), \eprint{1605.03781}.

\bibitem[{\citenamefont{Cai et~al.}(2020)\citenamefont{Cai, Pi, and
  Sasaki}}]{Cai:2019cdl}
\bibinfo{author}{\bibfnamefont{R.-G.} \bibnamefont{Cai}},
  \bibinfo{author}{\bibfnamefont{S.}~\bibnamefont{Pi}}, \bibnamefont{and}
  \bibinfo{author}{\bibfnamefont{M.}~\bibnamefont{Sasaki}},
  \bibinfo{journal}{Phys. Rev. D} \textbf{\bibinfo{volume}{102}},
  \bibinfo{pages}{083528} (\bibinfo{year}{2020}), \eprint{1909.13728}.

\bibitem[{\citenamefont{Hook et~al.}(2021)\citenamefont{Hook, Marques-Tavares,
  and Racco}}]{Hook:2020phx}
\bibinfo{author}{\bibfnamefont{A.}~\bibnamefont{Hook}},
  \bibinfo{author}{\bibfnamefont{G.}~\bibnamefont{Marques-Tavares}},
  \bibnamefont{and} \bibinfo{author}{\bibfnamefont{D.}~\bibnamefont{Racco}},
  \bibinfo{journal}{JHEP} \textbf{\bibinfo{volume}{02}}, \bibinfo{pages}{117}
  (\bibinfo{year}{2021}), \eprint{2010.03568}.

\bibitem[{\citenamefont{Ellis et~al.}(2014)\citenamefont{Ellis, Vallisneri,
  Taylor, and Baker}}]{Ellis:2020zenodo}
\bibinfo{author}{\bibfnamefont{J.~A.} \bibnamefont{Ellis}},
  \bibinfo{author}{\bibfnamefont{M.}~\bibnamefont{Vallisneri}},
  \bibinfo{author}{\bibfnamefont{S.~R.} \bibnamefont{Taylor}},
  \bibnamefont{and} \bibinfo{author}{\bibfnamefont{P.~T.} \bibnamefont{Baker}},
  \bibinfo{journal}{Zenodo}  (\bibinfo{year}{2014}).

\bibitem[{\citenamefont{S.R.~Taylor and Vigeland.}(2021)}]{extensions2021}
\bibinfo{author}{\bibfnamefont{J.~H. J.~S.} \bibnamefont{S.R.~Taylor},
  \bibfnamefont{P.T.~Baker}} \bibnamefont{and}
  \bibinfo{author}{\bibfnamefont{S.}~\bibnamefont{Vigeland.}},
  \bibinfo{howpublished}{\url{https://github.com/nanograv/enterprise_extensions}}
  (\bibinfo{year}{2021}).

\bibitem[{\citenamefont{Lamb et~al.}(2023)\citenamefont{Lamb, Taylor, and van
  Haasteren}}]{Lamb:2023jls}
\bibinfo{author}{\bibfnamefont{W.~G.} \bibnamefont{Lamb}},
  \bibinfo{author}{\bibfnamefont{S.~R.} \bibnamefont{Taylor}},
  \bibnamefont{and} \bibinfo{author}{\bibfnamefont{R.}~\bibnamefont{van
  Haasteren}} (\bibinfo{year}{2023}), \eprint{2303.15442}.

\bibitem[{\citenamefont{Mitridate et~al.}(2023)\citenamefont{Mitridate, Wright,
  von Eckardstein, Schr\"oder, Nay, Olum, Schmitz, and
  Trickle}}]{Mitridate:2023oar}
\bibinfo{author}{\bibfnamefont{A.}~\bibnamefont{Mitridate}},
  \bibinfo{author}{\bibfnamefont{D.}~\bibnamefont{Wright}},
  \bibinfo{author}{\bibfnamefont{R.}~\bibnamefont{von Eckardstein}},
  \bibinfo{author}{\bibfnamefont{T.}~\bibnamefont{Schr\"oder}},
  \bibinfo{author}{\bibfnamefont{J.}~\bibnamefont{Nay}},
  \bibinfo{author}{\bibfnamefont{K.}~\bibnamefont{Olum}},
  \bibinfo{author}{\bibfnamefont{K.}~\bibnamefont{Schmitz}}, \bibnamefont{and}
  \bibinfo{author}{\bibfnamefont{T.}~\bibnamefont{Trickle}}
  (\bibinfo{year}{2023}), \eprint{2306.16377}.

\bibitem[{\citenamefont{Ellis and van Haasteren}(2017)}]{Ellis:2017}
\bibinfo{author}{\bibfnamefont{J.}~\bibnamefont{Ellis}} \bibnamefont{and}
  \bibinfo{author}{\bibfnamefont{R.}~\bibnamefont{van Haasteren}},
  \bibinfo{howpublished}{\url{jellis18/ptmcmcsampler: Official release}}
  (\bibinfo{year}{2017}).

\bibitem[{\citenamefont{Kim}(1979)}]{Kim:1979if}
\bibinfo{author}{\bibfnamefont{J.~E.} \bibnamefont{Kim}},
  \bibinfo{journal}{Phys. Rev. Lett.} \textbf{\bibinfo{volume}{43}},
  \bibinfo{pages}{103} (\bibinfo{year}{1979}).

\bibitem[{\citenamefont{Shifman et~al.}(1980)\citenamefont{Shifman, Vainshtein,
  and Zakharov}}]{Shifman:1979if}
\bibinfo{author}{\bibfnamefont{M.~A.} \bibnamefont{Shifman}},
  \bibinfo{author}{\bibfnamefont{A.~I.} \bibnamefont{Vainshtein}},
  \bibnamefont{and} \bibinfo{author}{\bibfnamefont{V.~I.}
  \bibnamefont{Zakharov}}, \bibinfo{journal}{Nucl. Phys. B}
  \textbf{\bibinfo{volume}{166}}, \bibinfo{pages}{493} (\bibinfo{year}{1980}).

\bibitem[{\citenamefont{Dine et~al.}(1981)\citenamefont{Dine, Fischler, and
  Srednicki}}]{Dine:1981rt}
\bibinfo{author}{\bibfnamefont{M.}~\bibnamefont{Dine}},
  \bibinfo{author}{\bibfnamefont{W.}~\bibnamefont{Fischler}}, \bibnamefont{and}
  \bibinfo{author}{\bibfnamefont{M.}~\bibnamefont{Srednicki}},
  \bibinfo{journal}{Phys. Lett. B} \textbf{\bibinfo{volume}{104}},
  \bibinfo{pages}{199} (\bibinfo{year}{1981}).

\bibitem[{\citenamefont{Zhitnitsky}(1980)}]{Zhitnitsky:1980tq}
\bibinfo{author}{\bibfnamefont{A.~R.} \bibnamefont{Zhitnitsky}},
  \bibinfo{journal}{Sov. J. Nucl. Phys.} \textbf{\bibinfo{volume}{31}},
  \bibinfo{pages}{260} (\bibinfo{year}{1980}).

\bibitem[{\citenamefont{Li and Liu}(2003)}]{Li:2002xd}
\bibinfo{author}{\bibfnamefont{T.-j.} \bibnamefont{Li}} \bibnamefont{and}
  \bibinfo{author}{\bibfnamefont{T.}~\bibnamefont{Liu}}, \bibinfo{journal}{Eur.
  Phys. J. C} \textbf{\bibinfo{volume}{28}}, \bibinfo{pages}{545}
  (\bibinfo{year}{2003}), \eprint{hep-th/0204128}.

\bibitem[{\citenamefont{Kibble}(1976)}]{Kibble:1976sj}
\bibinfo{author}{\bibfnamefont{T.~W.~B.} \bibnamefont{Kibble}},
  \bibinfo{journal}{J. Phys. A} \textbf{\bibinfo{volume}{9}},
  \bibinfo{pages}{1387} (\bibinfo{year}{1976}).

\bibitem[{\citenamefont{Kibble}(1980)}]{Kibble:1980mv}
\bibinfo{author}{\bibfnamefont{T.~W.~B.} \bibnamefont{Kibble}},
  \bibinfo{journal}{Phys. Rept.} \textbf{\bibinfo{volume}{67}},
  \bibinfo{pages}{183} (\bibinfo{year}{1980}).

\bibitem[{\citenamefont{Vilenkin}(1981)}]{Vilenkin:1981iu}
\bibinfo{author}{\bibfnamefont{A.}~\bibnamefont{Vilenkin}},
  \bibinfo{journal}{Phys. Rev. Lett.} \textbf{\bibinfo{volume}{46}},
  \bibinfo{pages}{1169} (\bibinfo{year}{1981}), \bibinfo{note}{[Erratum:
  Phys.Rev.Lett. 46, 1496 (1981)]}.

\bibitem[{\citenamefont{Vilenkin and Everett}(1982)}]{Vilenkin:1982ks}
\bibinfo{author}{\bibfnamefont{A.}~\bibnamefont{Vilenkin}} \bibnamefont{and}
  \bibinfo{author}{\bibfnamefont{A.~E.} \bibnamefont{Everett}},
  \bibinfo{journal}{Phys. Rev. Lett.} \textbf{\bibinfo{volume}{48}},
  \bibinfo{pages}{1867} (\bibinfo{year}{1982}).

\bibitem[{\citenamefont{Kibble et~al.}(1982)\citenamefont{Kibble, Lazarides,
  and Shafi}}]{Kibble:1982dd}
\bibinfo{author}{\bibfnamefont{T.~W.~B.} \bibnamefont{Kibble}},
  \bibinfo{author}{\bibfnamefont{G.}~\bibnamefont{Lazarides}},
  \bibnamefont{and} \bibinfo{author}{\bibfnamefont{Q.}~\bibnamefont{Shafi}},
  \bibinfo{journal}{Phys. Rev. D} \textbf{\bibinfo{volume}{26}},
  \bibinfo{pages}{435} (\bibinfo{year}{1982}).

\bibitem[{\citenamefont{Preskill and Vilenkin}(1993)}]{Preskill:1992ck}
\bibinfo{author}{\bibfnamefont{J.}~\bibnamefont{Preskill}} \bibnamefont{and}
  \bibinfo{author}{\bibfnamefont{A.}~\bibnamefont{Vilenkin}},
  \bibinfo{journal}{Phys. Rev. D} \textbf{\bibinfo{volume}{47}},
  \bibinfo{pages}{2324} (\bibinfo{year}{1993}), \eprint{hep-ph/9209210}.

\bibitem[{\citenamefont{Martins
  et~al.}(2016{\natexlab{a}})\citenamefont{Martins, Rybak, Avgoustidis, and
  Shellard}}]{Martins:2016lzc}
\bibinfo{author}{\bibfnamefont{C.~J. A.~P.} \bibnamefont{Martins}},
  \bibinfo{author}{\bibfnamefont{I.~Y.} \bibnamefont{Rybak}},
  \bibinfo{author}{\bibfnamefont{A.}~\bibnamefont{Avgoustidis}},
  \bibnamefont{and} \bibinfo{author}{\bibfnamefont{E.~P.~S.}
  \bibnamefont{Shellard}}, \bibinfo{journal}{Phys. Rev. D}
  \textbf{\bibinfo{volume}{94}}, \bibinfo{pages}{116017}
  (\bibinfo{year}{2016}{\natexlab{a}}), \bibinfo{note}{[Erratum: Phys.Rev.D 95,
  039902 (2017)]}, \eprint{1612.08863}.

\bibitem[{\citenamefont{Martins
  et~al.}(2016{\natexlab{b}})\citenamefont{Martins, Rybak, Avgoustidis, and
  Shellard}}]{Martins:2016ois}
\bibinfo{author}{\bibfnamefont{C.~J. A.~P.} \bibnamefont{Martins}},
  \bibinfo{author}{\bibfnamefont{I.~Y.} \bibnamefont{Rybak}},
  \bibinfo{author}{\bibfnamefont{A.}~\bibnamefont{Avgoustidis}},
  \bibnamefont{and} \bibinfo{author}{\bibfnamefont{E.~P.~S.}
  \bibnamefont{Shellard}}, \bibinfo{journal}{Phys. Rev. D}
  \textbf{\bibinfo{volume}{93}}, \bibinfo{pages}{043534}
  (\bibinfo{year}{2016}{\natexlab{b}}), \eprint{1602.01322}.

\bibitem[{\citenamefont{Huang and Sikivie}(1985)}]{Huang:1985tt}
\bibinfo{author}{\bibfnamefont{M.~C.} \bibnamefont{Huang}} \bibnamefont{and}
  \bibinfo{author}{\bibfnamefont{P.}~\bibnamefont{Sikivie}},
  \bibinfo{journal}{Phys. Rev. D} \textbf{\bibinfo{volume}{32}},
  \bibinfo{pages}{1560} (\bibinfo{year}{1985}).

\bibitem[{\citenamefont{Blasi et~al.}(2023{\natexlab{b}})\citenamefont{Blasi,
  Mariotti, Rase, Sevrin, and Turbang}}]{Blasi:2022ayo}
\bibinfo{author}{\bibfnamefont{S.}~\bibnamefont{Blasi}},
  \bibinfo{author}{\bibfnamefont{A.}~\bibnamefont{Mariotti}},
  \bibinfo{author}{\bibfnamefont{A.}~\bibnamefont{Rase}},
  \bibinfo{author}{\bibfnamefont{A.}~\bibnamefont{Sevrin}}, \bibnamefont{and}
  \bibinfo{author}{\bibfnamefont{K.}~\bibnamefont{Turbang}},
  \bibinfo{journal}{JCAP} \textbf{\bibinfo{volume}{04}}, \bibinfo{pages}{008}
  (\bibinfo{year}{2023}{\natexlab{b}}), \eprint{2210.14246}.

\bibitem[{\citenamefont{Hiramatsu et~al.}(2011)\citenamefont{Hiramatsu,
  Kawasaki, Sekiguchi, Yamaguchi, and Yokoyama}}]{Hiramatsu:2010yu}
\bibinfo{author}{\bibfnamefont{T.}~\bibnamefont{Hiramatsu}},
  \bibinfo{author}{\bibfnamefont{M.}~\bibnamefont{Kawasaki}},
  \bibinfo{author}{\bibfnamefont{T.}~\bibnamefont{Sekiguchi}},
  \bibinfo{author}{\bibfnamefont{M.}~\bibnamefont{Yamaguchi}},
  \bibnamefont{and} \bibinfo{author}{\bibfnamefont{J.}~\bibnamefont{Yokoyama}},
  \bibinfo{journal}{Phys. Rev. D} \textbf{\bibinfo{volume}{83}},
  \bibinfo{pages}{123531} (\bibinfo{year}{2011}), \eprint{1012.5502}.

\bibitem[{\citenamefont{Lewis}(2019)}]{Lewis:2019xzd}
\bibinfo{author}{\bibfnamefont{A.}~\bibnamefont{Lewis}} (\bibinfo{year}{2019}),
  \eprint{1910.13970}.

\bibitem[{\citenamefont{Gouttenoire and
  Vitagliano}(2023)}]{Gouttenoire:2023ftk}
\bibinfo{author}{\bibfnamefont{Y.}~\bibnamefont{Gouttenoire}} \bibnamefont{and}
  \bibinfo{author}{\bibfnamefont{E.}~\bibnamefont{Vitagliano}}
  (\bibinfo{year}{2023}), \eprint{2306.17841}.

\end{thebibliography}

\clearpage
\newpage
\maketitle
\onecolumngrid
\begin{center}
\textbf{\large Clockwork axion footprint on nano-hertz stochastic gravitational wave background} \\ 
\vspace{0.05in}
{ \it \large Supplemental Material}\\ 
{By Bo-Qiang Lu, Cheng-Wei Chiang, and Tianjun Li}
\vspace{0.05in}
\end{center}
\onecolumngrid
\setcounter{equation}{0}
\setcounter{figure}{0}
\setcounter{table}{0}
\setcounter{section}{0}
\setcounter{page}{1}
\makeatletter
\renewcommand{\theequation}{S\arabic{equation}}
\renewcommand{\thefigure}{S\arabic{figure}}
\renewcommand{\thetable}{S\arabic{table}}

In this supplemental material, we provide detailed calculations and analysis results for our work.

\section{Mass eigenstates}

Following the paradigm of the clockwork axion framework~\cite{Kaplan:2015fuy}, we introduce $N+1$ copies of complex scalars, denoted by $\Phi_j(x)$ with $j=0,1,...,N$. The potential of these scalars is given by
\begin{equation}
    \label{eq:poten1}
    V(\Phi)=\sum_{j=0}^{N}\left(-m^{2}\left|\Phi_{j}\right|^{2}+\lambda\left|\Phi_{j}\right|^{4}\right)-
    \varepsilon \sum_{j=0}^{N-1}\left( \Phi_{j}^{\dagger} \Phi_{j+1}^{3}+ {\rm H.c.} \right),
\end{equation}
where $m^2$, $\lambda$, and $\varepsilon$ have been assumed to be real and universal for the scalars.  
The first term respects a global $U(1)^{N+1}$ symmetry, $\Phi_{j} \rightarrow \exp \left[i\theta_j\right] \Phi_{j}$, which is explicitly broken by the $\varepsilon$-dependent terms down to a global $U(1)$ symmetry, 
$\Phi_{j} \rightarrow \exp \left[i q^{N-j} \theta\right] \Phi_{j}$, with $\theta \in [0, 2\pi)$ and $q\equiv 3$.  This exact $U(1)$ is identified as the PQ symmetry.
The spontaneous breakdown of $U(1)^{N+1}$ takes place when the radial components acquire a vacuum expectation value $\left \langle \Phi_j \right \rangle=f/\sqrt{2}$ with $f=m / \sqrt{\lambda }$,
which results in $N$ massive pseudo-Goldstone bosons $A_i$ and one massless Goldstone boson $a$.

After the spontaneous symmetry breaking, we parametrize the scalar fields as $\Phi_i=(f+\rho_i)e^{i\pi_i/f}/\sqrt{2}$. 
The mass for the radial component $\rho_i$ is $m_{\rho}=\sqrt{2\lambda}f$.
The potential for the $N+1$ Goldstone bosons is given by
\begin{equation}
    \begin{aligned}
    \label{eq:Vpi}
    V(\pi)&=\frac{1}{4}\varepsilon f^4 \sum_{j=0}^{N-1} e^{i\left(q \pi_{j+1}-\pi_j\right)/f}+{\rm H.c.}+\cdots\\
    &=\frac{1}{2} \varepsilon f^{4} \sum_{j=0}^{N-1} \cos \frac{q \pi_{j+1}-\pi_{j}}{f} 
    \simeq \frac{1}{2}\sum_{i,j=0}^{N}\pi_j \left( M_{\pi}^2 \right)_{ji} \pi_i,
    \end{aligned}
\end{equation}        
where the constant term is omitted and the mass matrix
\begin{equation}
    M_{\pi}^{2}=m_{G}^{2}\left(\begin{array}{cccccc}
    1 & -q & 0 & \cdots & & 0 \\
    -q & 1+q^{2} & -q & \cdots & & 0 \\
    0 & -q & 1+q^{2} & \cdots & & 0 \\
    \vdots & \vdots & \vdots & \ddots & & \vdots \\
    & & & & 1+q^{2} & -q \\
    0 & 0 & 0 & \cdots & -q & q^{2}
    \end{array}\right),
    \label{Mass-Matrix}
\end{equation}
where $m_G^2=\varepsilon f^2/2$. One then rotates the $\pi_i$ fields to the mass eigenstates $a_i \equiv (a, A_1, \dots, A_N)$ by a real $(N+1)\times (N+1)$ orthogonal matrix $O$, so that the mass matrix is diagonalized as
$O^{T} M_{\pi}^{2} O=\operatorname{diag}\left(m_{a}^{2},m_{A_1}^2, \ldots, m_{A_{N}}^{2}\right)$, 
where the eigenvalues of $N+1$ Goldstone bosons $a_i$ are given by
\begin{equation}
    \label{eq:massG}
    m_{a}^2=0~{\rm and}~m_{A_k}^2=\eta_km_{G}^2,
\end{equation} 
with $\eta_{k}\equiv q^{2}+1-2q\cos\frac{k\pi}{N+1}~\left( k=1,2,...,N \right)$.

The massless Goldstone boson $a$ is identified as the QCD axion and the $N$ massive pseudo-Goldstone states $A_k$ are the so-called gear 
fields since they play the role of `gears' in the clockwork mechanism.
We adopt the approximate mass relation $m_{A_k}^2\simeq m_{A}^2\simeq \varepsilon f^2$ for the massive axions $A_k$.
The matrix elements of $O$ are given by
\begin{equation}
    \label{eq:rote0}
    O_{i 0}=\frac{\mathcal{N}_{0}}{q^{i}}, \quad O_{i k}=\mathcal{N}_{k}\left[q \sin \frac{i k \pi}{N+1}-\sin \frac{(i+1) k \pi}{N+1}\right],
\end{equation}
with $i=0,1,...,N$, $k=1,2,...,N$, and
\begin{equation}
    \label{eq:rote1}
    \mathcal{N}_{0} \equiv \sqrt{\frac{q^{2}-1}{q^{2}-q^{-2 N}}}
    ~, \quad 
    \mathcal{N}_{k} \equiv \sqrt{\frac{2}{(N+1) \eta_{k}}}
    ~.
\end{equation}

The $(N+1)$ $a_i$ fields are related to the $\pi_i$ fields by the rotation
\begin{equation}
    \label{eq:piN}
    \pi_i=\sum_{j=0}^{N}O_{ij}a_{j}
    \equiv O_{i0}a+\sum_{j=1}^{N}O_{ij}A_j
    ~.
\end{equation}
The potential of the (pseudo-)Goldstone bosons in the physical basis is then given by the sum of the contributions from all the sites
\begin{equation}
    \label{eq:gearV}
    V(\pi)=\sum_{j=0}^{N} V_j(A_j)=\frac{1}{2}m_{G}^2\sum_{j=1}^{N}\eta_jA_j^2=\frac{1}{4}\varepsilon f^2\sum_{j=1}^{N}\eta_jA_j^2
    ~.
\end{equation}
Here we have used the fact that $m_a^2=0$.

The clockwork mechanism is illustrated as follows.  Consider the effective Lagrangian in which the $N$-th site $\pi_N$ is coupled 
to the QCD topological term
\begin{equation}
    \label{eq:qcdtopol}
    \mathcal{L}\supset\frac{\alpha_{s}}{8 \pi}\frac{\pi_{N}}{f} G_{\mu \nu}^{a} \tilde{G}^{\mu \nu, a}
    ~,
\end{equation}
where $\alpha_s=g_s^2/(4\pi)$ is the strong coupling constant which runs with the energy scale and $G_{\mu \nu}^{a}$ is the gluon field strength tensor. 
This gluon anomalous term can be fulfilled in the Kim-Shifman-Vainshtein-Zakharov (KSVZ)~\cite{Kim:1979if,Shifman:1979if} type or the 
Dine-Fischler-Srednicki-Zhitnitsky (DFSZ)~\cite{Dine:1981rt,Zhitnitsky:1980tq} type of axion models. Note that the DFSZ type axion has a DW number
$N_{\rm DW}>1$ after the QCD phase transition.  Yet we restrict ourselves in the case with one DW; otherwise, we need to introduce another bias potential. Using Eq.~\eqref{eq:piN}, the axion coupling to the topological term is then given by
\begin{equation}
    \mathcal{L}\supset\frac{\alpha_{s}}{8 \pi}\frac{a}{f_{a}} G_{\mu \nu}^{a} \tilde{G}^{\mu \nu, a}
    ~,
\end{equation}
where we have defined
\begin{equation}
    \label{eq:decayCont}
    f_a \equiv \frac{f}{O_{N0}}=\frac{q^Nf}{\mathcal{N}_{0}}\simeq q^Nf
    ~.
\end{equation} 
If the QCD topological term~\eqref{eq:qcdtopol} occurs at the `first' site $i=N$, we observe from Eq.~\eqref{eq:rote0} that the coupling of the massless axion at the `last' site $i=0$ is suppressed by a factor of $q^N$.  
In other words, the axion decay constant $f_a$ is amplified by a factor of $q^N$ compared to the symmetry-breaking scale $f$, as given in Eq.~\eqref{eq:decayCont}.

Furthermore, the UV completion of our model could be a five-dimensional orbifold $U(1)_{PQ}$ model on
the space-time $M^4\times S^1/(Z_2\times Z_2')$~\cite{Li:2002xd}. In particular, for $q=1$, the mass matrix in Eq.~\eqref{Mass-Matrix} 
is the same as the mass matrix in Eq.~(36) in Ref.~\cite{Li:2002xd}, where the corresponding
eigenvalues and eigenvectors are given in Eqs.~($37-40$) as well. The UV completion will be studied in detail elsewhere.

\section{String-wall evolution in VOS model}
\label{sec:VOSwithfric}

For small values of $\varepsilon$, we can treat the explicit symmetry-breaking term as a perturbation, with the $U(1)^{N+1}$ being exact at high temperatures. The spontaneous $U(1)^{N+1}$  
symmetry-breaking produces $N+1$ angular degrees of freedom $\pi_i$ ($i=0,...,N$) with a shift symmetry $\pi_i\to \pi_i+{\rm const}$. There appear $N+1$ kinds of cosmic $\pi$-strings corresponding to the topological configurations of $\pi_{i}$'s. The $\pi$-string with a core size $\lambda_{\rho}\simeq 1/(\sqrt{2\lambda}f)$ has a tension~\cite{Kibble:1976sj,Kibble:1980mv,Vilenkin:1981iu,Vilenkin:1982ks} 
\begin{equation}
    \mu \simeq \pi f^2+\pi f^2 \log \left(t/\lambda_{\rho}\right).
\end{equation}
Subsequently, the $\varepsilon$-dependent terms become important and break the $N$ global $U(1)$ symmetries down to their discrete subgroups. 
DWs are formed with a tension of $\sigma\simeq 8m_{A}f^2$, where $m_A\simeq \varepsilon^{1/2}f$. 
The $\pi$-strings form the boundaries of the walls and get connected to the walls~\cite{Kibble:1982dd}.
A single string can be stretched between a number of walls and form a complicated string-wall structure. 
The $a$-string associated with the spontaneous breakdown of the exact PQ $U(1)$ symmetry is topologically identified 
as an isolated string bundle, which however is found not to appear in the evolution of the network with $N\ge 3$~\cite{Long:2018nsl}. 
After the formation, DW's tension immediately dominates the dynamics of the string-wall network. 

The string-wall network that consists of a large number of $\pi$-strings connected by more than two DWs is stable~\cite{Higaki:2016jjh,Hiramatsu:2012sc} since the string stretched by one wall can also be stretched by another.
However, the walls can be destroyed locally by the strings bound to the walls by creating holes in the walls.
The nucleation rate of a hole via quantum and thermal fluctuations is approximated as
$\Gamma_S \sim \sigma e^{-S_{\rm E}}$~\cite{Kibble:1982dd,Preskill:1992ck},
where $S_{\rm E}=4\pi R^2\mu - \frac{4}{3}\pi R^3 \sigma$
is the spherical Euclidean action of the bounce solution.
In the semiclassical approximation, the tunneling proceeds via the stationary path given by
$S_0=16 \pi \mu^3/(3 \sigma^2)$ at $R_0=2\mu/\sigma$ by the minimization of $S_{\rm E}$. 
We find that with $\varepsilon\lesssim 1$, $\mu^3/\sigma^2\gtrsim 1$ is always fulfilled to suppress the hole nucleation.

The cosmological evolution of the topological defect network with a characteristic length scale $L$ and a root-mean-squared velocity $v$ 
can be quantitatively described by the velocity-dependent one-scale (VOS) model~\cite{Martins:2016lzc}
\begin{equation}
    (4-n) \frac{d L}{d t}=(4-n) H L+v^2 \frac{L}{\ell_d}+c v
    ~,
\end{equation}
\begin{equation}
    \frac{dv}{dt}=\left( 1-v^2\right)\left( \frac{k}{L}-\frac{v}{\ell_d}\right)
    ~,
\end{equation}
where $n$ is the dimension of the defect worldsheet, with $n=1,~2,~3$ for monopoles, cosmic strings and DWs, respectively.
In the radiation dominant Universe, we take $c=0.81\pm 0.04$ and $k= 0.66\pm 0.04$ for the DW according to numerical simulations~\cite{Martins:2016ois}. 
The characteristic damping length scale
\begin{equation}
    \frac{1}{\ell_d}=nH+\frac{1}{\ell_f} 
\end{equation}
includes the damping effects on the network oscillation from Hubble drag and particle friction.

We summarize the characteristic time scales for the evolution of the string-wall network in Table~\ref{tab:timescale}.
As shown in the table, for the string-wall network considered in this work, the time scale for DWs to become dominant is
$t_d\sim 10^{-4}~\rm GeV^{-1}$, which is much shorter than the other time scales. Therefore, as the DW forms, the evolution of the string-wall network
is immediately taken over by the wall tension. Furthermore, from Table~\ref{tab:timescale}, we observe that the string and wall 
are formed nearly at the same time. Thus, the string-wall network is controlled by the wall tension from the beginning. 
Therefore we focus on the DW evolution in the VOS model. The characteristic length scale $L$ is related to the wall energy density $\rho_w$ via the one-scale ansatz
\begin{equation}
    \rho_{w}=\frac{\sigma}{L}.
\end{equation}
The friction length $\ell_f$ is determined by the reflection pressure $P$ by
\begin{equation}
    \frac{1}{\ell_f}\simeq \frac{P}{\sigma v}.
\end{equation}
Note that the reflectivity $\zeta$ is in general momentum-dependent. Suppose the Goldstone bosons couple to photons or gluons via the anomaly.
By solving the Schr$\ddot{\rm o}$dinger equation, Ref.~\cite{Huang:1985tt} shows that the reflection rate $\zeta$ of a relativistic particle 
by the DW depends on both the particle's injection momentum $p_z$ and the width of the wall $m_{A}^{-1}$. 
For the particles with $p_z\gg m_{A}$ or $p_z\ll m_{A}$, the reflection is strongly suppressed, i.e., $\zeta\simeq 0$. 
On the other hand, the reflection $\zeta\simeq \alpha^2$ when $p_z\sim m_{A}$~\cite{Huang:1985tt}.
This in general leads to an exponential suppression $e^{-m_{A}/T}$ on the reflection pressure for $T\ll m_{A}$ and, thus,
the friction effects from the surrounding plasma become negligible.
Therefore, we expect that the plasma friction could be significant at temperature $ 0.1m_{A}\lesssim T\lesssim m_{A}$ and 
limit the growth of the wall~\cite{Blasi:2022ayo}. The reflection pressure can be estimated as
\begin{eqnarray}\label{eq:fricpress}
    P\simeq 
    	\begin{cases}
        \displaystyle
        \frac{\alpha^2}{\pi^2}m_A^2T^2\quad\quad\quad\quad {\rm for}\quad { T\gtrsim 0.1m_A}, \\ 
        \displaystyle
        \frac{\alpha^2}{\pi^2}m_A^3Te^{-m_A/T}\quad {\rm for}\quad { T\lesssim 0.1m_A}. 
        \end{cases}
\end{eqnarray}
where $\alpha=\alpha_s$ if the massive axion $A$ is coupled to gluons via the topological term~\eqref{eq:qcdtopol}.

\begin{table}[tbp]
  \centering
  \caption{\label{tab:timescale} Time table for the evolution of string-wall network}
  \begin{threeparttable}
  \renewcommand\arraystretch{1.5}
  \begin{tabular}{|c||c|}
  \hline
  $\quad$ \tnote{a}~~Time $\quad$ & $\quad$ Event $\quad$ \\
  \hline 
  $\quad$ $t_s=M_{\rm pl}/(3.32g_*^{1/2}f^2)\sim 9\times 10^{6}~{\rm GeV^{-1}}$ $\quad$ & $\quad$ $N+1$ $U(1)$ breaking, string formation $\quad$ \\
  \hline 
  $\quad$ $t_w=M_{\rm pl}/(3.32g_*^{1/2}\varepsilon f^2)\sim 9\times 10^{6}/\varepsilon~{\rm GeV^{-1}}$ $\quad$ & $\quad$ $N$ discrete symmetries breaking, DW formation $\quad$ \\
  \hline
  $\quad$ $t_c=\mu/\sigma\sim 5\times 10^{-5}/\varepsilon^{1/2}~{\rm GeV^{-1}}$ $\quad$ & $\quad$ DW tension dominates network dynamics $\quad$ \\
  \hline
  $\quad$ \tnote{b}~~$t_{\rm QCD}=M_{\rm pl}/(3.32g_*^{1/2}T_{\rm QCD}^2)\sim (0.1-1)\times 10^{19}~{\rm GeV^{-1}}$ $\quad$ & $\quad$ QCD phase transition, bias potential generate $\quad$ \\ 
  \hline
  $\quad$ \tnote{c}~~$t_{\rm ann}=\sigma/(2cV_b)\sim \varepsilon^{1/2}\times 10^{19}~{\rm GeV^{-1}}$ $\quad$ & $\quad$ DWs begin annihilation $\quad$ \\
  \hline
  $\quad$ \tnote{d}~~$t_{\rm PBH}\gtrsim 1/(10\pi G \sigma)\sim 5\times 10^{19}/\varepsilon^{1/2}~{\rm GeV^{-1}}$ $\quad$ & $\quad$ PBHs form significantly $\quad$ \\
  \hline
  $\quad$ $t_{\rm dom}=3M_{\rm pl}^2/(32\pi \sigma)\sim 7\times 10^{19}/\varepsilon^{1/2}~{\rm GeV^{-1}}$ $\quad$ & $\quad$ DWs dominate the Universe density $\quad$ \\
  \hline 
  $\quad$ $t_*=M_{\rm pl}^2/(\mu m_{\rho})\sim 2\times 10^{20}~{\rm GeV^{-1}}$ $\quad$ & $\quad$ String damping time catchs up with its size $\quad$ \\
  \hline
  \end{tabular}
  \begin{tablenotes}
    \footnotesize
    \item{\bf Notes.}
    \item{a:} We adopt $f\simeq 200$~TeV, $1~{\rm GeV^{-1}}=6.58\times 10^{-25}$~s.
    \item{b:} Uncertainty exists in $T_{\rm QCD}$.
    \item{c:} We adopt the condition $c\sigma H\simeq V_b$, where $c\sim \mathcal{O}(1)$.
    \item{d:} PBHs do not form if $t_{\rm PBH}>t_{\rm ann}$.
  \end{tablenotes}
  \end{threeparttable}
\end{table} 

\begin{figure}
    \centering
    \includegraphics[width=120mm,angle=0]{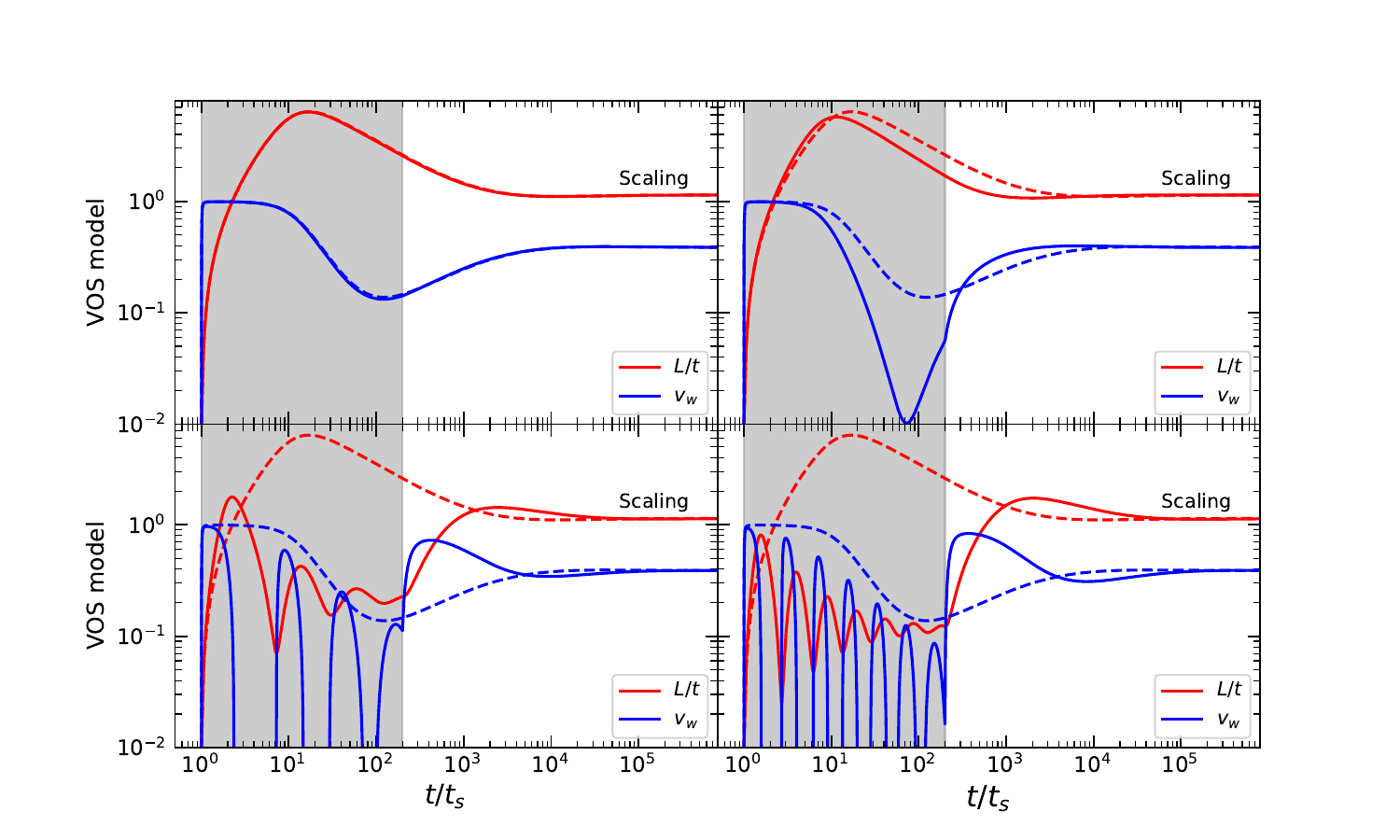}
    \caption{String-wall network evolution with cosmic time $t/t_s$. We take $f=200$~TeV and $\varepsilon=0.5$.  The solid curves represent the 
    results with $\alpha=\alpha_s,~5\alpha_s,~18\alpha_s$, and $25\alpha_s$ from upper left to lower right, respectively. The dashed curves denote
    the frictionless evolution of the network, i.e., setting $\alpha=0$. The ratio of the network scale $L$ to the cosmic time $t$ and the wall
    velocity are represented by the red and blue curves, respectively. The temperature $T\gtrsim 0.1m_A$ in the grey region, where the friction effect could
    be most significant. We adopt $\alpha_s=1$ for illustration and note that $\alpha_s\lesssim 0.1$ at the energy scale of $\sim 100$~TeV. }
    \label{fig:friction}
\end{figure}

In Fig.~\ref{fig:friction}, the friction effect from the reflection of relativistic particles is depicted along with those with the friction turned off. We observe that the friction effect can be ignored for the massive axion $A$ whose
coupling to gluons $\alpha=\alpha_s$, and it becomes important when $\alpha$ is several times that of $\alpha_s$ (this may be achieved by introducing a number of quark flavors that are charged under the PQ $U(1)$ symmetry). The wall quickly becomes ultra-relativistic after its formation.  Its size continuously grows to exceed the Hubble horizon and reaches the maximum value of $L\simeq 7t$ at $t\simeq 10t_s$.
Then the frictions from the cosmic expansion and plasma reflection slow down its growth and eventually lead to shrinking. 
The wall size scales as $L\simeq t$ after entering the scaling regime at $t\simeq 10^{4}t_s$.
If the particle friction is sufficiently strong, the network rapidly oscillates and will decay into strong GWs and particle radiation, which however are not included in our VOS model. For the network considered here, the plasma friction is negligible. 

The total energy density of the string-wall network $\rho_{\rm sw}\simeq \rho_{s}+\rho_{w}$ in the scaling regime is given by
\begin{equation}
    \label{eq:rhowall}
    \rho_{w}(t)=\mathcal{A}\frac{\sigma}{t}
    ~~{\rm and}~~
    \rho_{s}(t)=\frac{\xi \mu}{t^2},
\end{equation}
where $\mathcal{A}\simeq 0.8\pm0.1$~\cite{Hiramatsu:2012sc} and $\xi=0.87\pm0.14$~\cite{Hiramatsu:2010yu} from simulations, and the cosmic time $t=1/(2H)$.   The Hubble expansion rate 
$H=1.66g_*^{1/2} T^2/M_{\mathrm{pl}}$, with $g_{*}$ being the effective relativistic degrees of freedom and the Planck scale $M_{\mathrm{pl}}=1.22 \times 10^{19}~ \mathrm{GeV}$. 
Eq.~\eqref{eq:rhowall} shows that the energy density of DWs in the scaling regime decreases as $\propto t^{-1}$, slower than that of strings 
and radiation, which are found to be $\propto t^{-2}$.
Therefore, when $\rho_{w}(t_{\rm dom})=\rho_c(t_{\rm dom}) \equiv 3H_{\rm dom}^2/(8\pi G)$, the energy density would be dominated
by the DWs at 
\begin{equation}
T_{\rm dom}=  5.44 \times 10^{-2} \mathrm{GeV} \varepsilon^{1 / 4}\left(\frac{g_*\left(T_{\rm dom}\right)}{10}\right)^{-1/4}\left(\frac{f}{100~\mathrm{TeV}}\right)^{3/2}.
\end{equation}
The temperature at which DWs annihilate significantly is given by Eq.~\eqref{eq:Tann}. We require that the DWs annihilate before they dominate the energy density of the Universe, i.e., $T_{\rm ann} \gtrsim T_{\rm dom}$, and obtain the constraint
\begin{equation}
f\lesssim 100 \mathrm{TeV} \varepsilon^{-1 / 6}\left( \frac{\Lambda_{\mathrm{QCD}}}{100~\mathrm{MeV}} \right)^{2 / 3} .
\end{equation}

\section{Details of the data analysis}
\label{sec:analysisdata}

The PTA searches for the GW signal by the timing-residual cross-power spectral density
\begin{equation}
    S_{a b}(\nu)=\Gamma_{a b} h_c^2(\nu) /\left(12 \pi^2 \nu^3\right),
\end{equation}
where $\Gamma_{a b}$ describes the average correlations between pulsars $a$ and $b$, and is given by the Hellings-Downs function for an isotropic and unpolarized GW background (GWB). The characteristic strain $h_c(\nu)$ can be used to determine the GWB relic abundance today via $\Omega _{\mathrm{GW}}(\nu)=2 \pi^2 \nu^2 h_c^2(\nu)/(3 H_0^2)$ where the Hubble constant is $H_0=100 h~\mathrm{km~s}^{-1}~\mathrm{Mpc}^{-1}$, 
with $h=0.67$ from the Planck observation~\cite{Planck:2018vyg}.

We implement the GW signal and PTA likelihood using the publicly available packages \texttt{enterprise}~\cite{Ellis:2020zenodo}, 
\texttt{enterprise\_extensions}~\cite{extensions2021}, \texttt{ceffyl}~\cite{Lamb:2023jls}, which are also encoded in 
\texttt{PTArcade}~\cite{Mitridate:2023oar}.
The Markov chain Monte Carlo (MCMC) tools are implemented in the \texttt{PTMCMCSampler} package~\cite{Ellis:2017} to 
sample the parameter points from the posterior distribution.
Following the searches for the GWB with a common power-law spectrum carried out by the IPTA~\cite{Antoniadis:2022pcn} 
and NANOGrav~\cite{NANOGrav:2023gor} Collaborations, only the first 14 frequencies of each dataset are included in the analysis.
In modeling the timing residuals, we take into account the white noise and red noise, including the pulsar-intrinsic red noise and common 
red noise produced by a GWB. 
We generate $2\times 10^6$ MC sample points for each analysis and model to fit each dataset separately. 
The parameter posterior distributions are derived using \texttt{GetDist}~\cite{Lewis:2019xzd}.
We provide the priors for noise and signal parameters in Table~\ref{tab:priors}, while the posteriors from the analysis are presented 
in Table~\ref{tab:posteriors}.

\begin{table}[tbp]
    \centering
    \caption{\label{tab:priors} Summary of model parameters and prior ranges.}
    \begin{threeparttable}
    \renewcommand\arraystretch{1.0}
    \begin{tabular}{llll}
    \hline
    \hline
    Parameter $\quad\quad$ & Description $\quad\quad$ & Prior $\quad\quad$ &  Comments \\
    \hline
    {\bf White noise } & & & \\
    $E_k$ $\quad\quad$ & EFAC per backend/receiver system $\quad\quad$ & Uniform $[0,~10]$ $\quad\quad$ & Single-pulsar analysis only \\
    $Q_k$[s] $\quad\quad$ & EQUAD per backend/receiver system $\quad\quad$ & Log-uniform $[-8.5,~-5]$ $\quad\quad$ & Single-pulsar analysis only \\
    $J_k$ [s] $\quad\quad$ & ECORR per backend/receiver system $\quad\quad$ & Log-uniform $[-8.5,~-5]$ $\quad\quad$ & Single-pulsar analysis only \\
    \hline
    {\bf Red noise } & & & \\
    $A_{\rm red}$ $\quad\quad$ & Red noise power-law amplitude $\quad\quad$ & Log-uniform $[-20,~-11]$ $\quad\quad$ & one parameter per pulsar \\
    $\gamma_{\rm red}$ $\quad\quad$ & Red noise power-law spectral index $\quad\quad$ & Uniform $[0,~7]$ $\quad\quad$ & one parameter per pulsar \\
    \hline
    {\bf GWB } & & & \\
    $\log_{10}A_{*}$ $\quad\quad$ & GWB power-law amplitude $\quad\quad$ & Log-uniform $[-18, -11]$ $\quad\quad$ & one parameter per PTA \\
    $\gamma_{*}$ $\quad\quad$ & GWB power-law spectral index $\quad\quad$ & Uniform $[0,~7]$ $\quad\quad$ & one parameter per PTA \\
    \hline
    {\bf DW } & & & \\
    $f$~[100~TeV] $\quad\quad$ & $U(1)$ breaking scale in unit 100~TeV $\quad\quad$ & Uniform $[0.5,~3.5]$ $\quad\quad$ & one parameter per PTA \\
    $\varepsilon$ $\quad\quad$ & Explicitly breaks $U(1)$ to discrete subgroup $\quad\quad$ & Uniform $[0.1,~1.0]$ $\quad\quad$ & one parameter per PTA \\
    $N$ $\quad\quad$ & Folds of $U(1)$ symmetry $\quad\quad$ & Uniform $[5,~20]$ $\quad\quad$ & one parameter per PTA \\
    \hline
    \hline
    \end{tabular}
    \end{threeparttable}
  \end{table} 
\begin{table}[tbp]
    \centering
    \caption{\label{tab:posteriors} Bayes Estimator for the fit to IPTA-DR2 and NG15 datasets with DW and DW+GWB.}
    \begin{threeparttable}
    \renewcommand\arraystretch{1.0}
    \begin{tabular}{lll}
    \hline
    \hline
    & \makecell[c]{{\bf IPTA-DR2}} $\quad\quad$ & \makecell[c]{{\bf NG15}} $\quad\quad$   \\
    \centering
    Parameter $\quad$ & $\quad\quad$\makecell[c]{DW $\quad\quad$ DW+GWB $\quad\quad$} & $\quad\quad\quad\quad$\makecell[c]{DW $\quad\quad$ DW+GWB $\quad\quad$} \\         
    \hline
    $f$~[100~TeV] $\quad\quad$ & \makecell[c]{1.76$\pm$0.22$\quad\quad$ 1.13$\pm$0.39 $\quad\quad$} & \makecell[c]{1.81$\pm$0.21$\quad\quad$ 1.72$\pm$0.25} \\
    $\varepsilon$ $\quad\quad$ & \makecell[c]{0.48$\pm$0.26$\quad\quad$ 0.46$\pm$0.25 $\quad\quad$} & \makecell[c]{0.50$\pm$0.26$\quad\quad$ 0.50$\pm$0.25}  \\
    $N$ $\quad\quad$ & \makecell[c]{12.28$\pm$4.35$\quad\quad$12.14$\pm$4.34 $\quad\quad$} & \makecell[c]{12.21$\pm$4.35$\quad\quad$ 12.33$\pm$4.32}  \\
    $\log_{10}A_{*}$ $\quad\quad$ & \makecell[c]{$\quad\quad\quad$...$\quad\quad\quad$ $-14.47\pm 0.14$ $\quad\quad$} & \makecell[c]{$\quad\quad\quad$...$\quad\quad\quad$ $-15.15\pm 0.40$}  \\
    $\gamma_{*}$ $\quad\quad$ & \makecell[c]{$~\quad\quad$...$\quad\quad\quad\quad$ 4.27$\pm$0.18 $\quad\quad$} & \makecell[c]{$~\quad\quad$...$\quad\quad\quad\quad$ 4.64$\pm$0.35}  \\
    \hline
    \hline
    \end{tabular} 
    \end{threeparttable}
  \end{table} 

\begin{figure}
    \centering
    \includegraphics[width=75mm,angle=0]{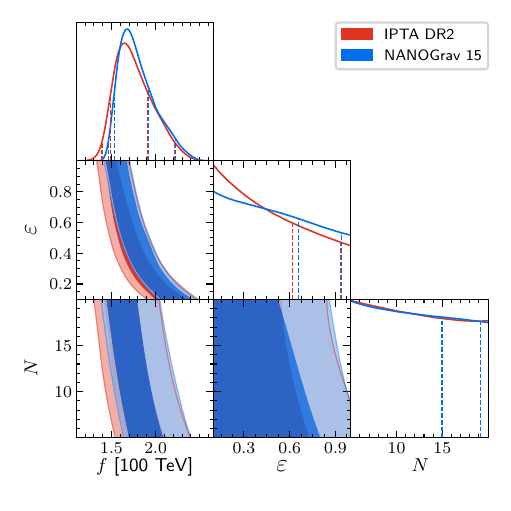}
    \includegraphics[width=75mm,angle=0]{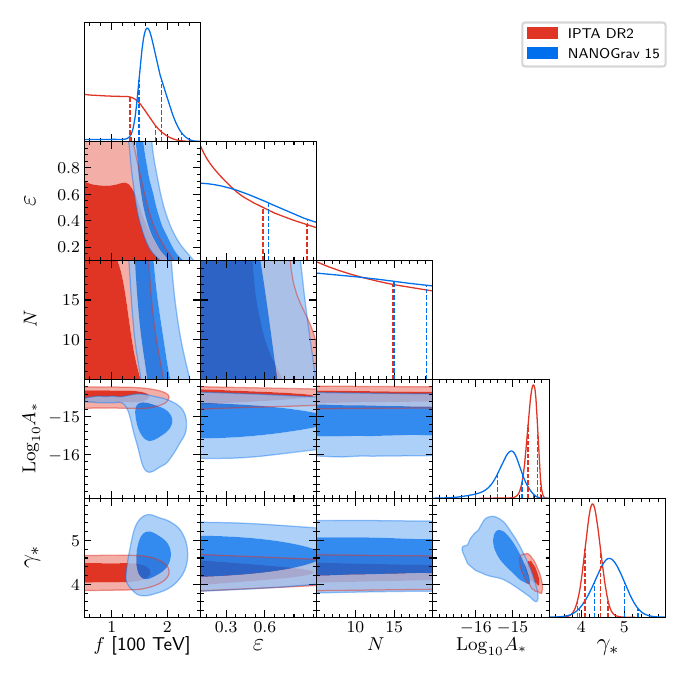}
    \caption{Corner plot of the posterior distributions of the parameters.}
    \label{fig:conrer}
\end{figure}

In order to investigate the possible effect on the DW interpretation for the PTA observations from the potential astrophysical GW sources, we also consider including a GWB source represented by a common power-law spectrum
\begin{equation}
    S_{*}(f)=\frac{A_{*}}{12 \pi^2}\left(\frac{f}{y r^{-1}}\right)^{-\gamma_{*}} \rm yr^3.
\end{equation}
The power spectral density (PSD) is related to the parameter $\Phi(f)$ by $\Phi(f)=S_*(f)/T_{\rm obs}$. 
For the GW emission induced by the orbital rotation of a population of SMBHB, the expected spectral index is $\gamma_{\rm BHB}=13/3$.
However, $A_{\rm BHB}$ is weakly constrained by the current observations and numerical simulations. 
We take the priors $[-18,~-11]$ and $[0,~7]$ for $\log_{10}A_{*}$ and $\gamma_{*}$, respectively.

Our main fit results are presented in Table~\ref{tab:posteriors} and Fig.~\ref{fig:conrer}. We observe from the left plot of 
Fig.~\ref{fig:conrer} that a single DW source can fit both datasets quite well. When a power-law-spectrum GWB source is included in the fit, the
IPTA-DR2 data tend to support the SMBHB origin since the $\gamma_*$ posterior is $4.27\pm 0.18$, which is in agreement with the spectrum index 
induced by SMBHB. On the other hand, the NG15 data supports the DW interpretation for the stochastic GWB signal.

\begin{figure}
    \centering
    \includegraphics[width=75mm,angle=0]{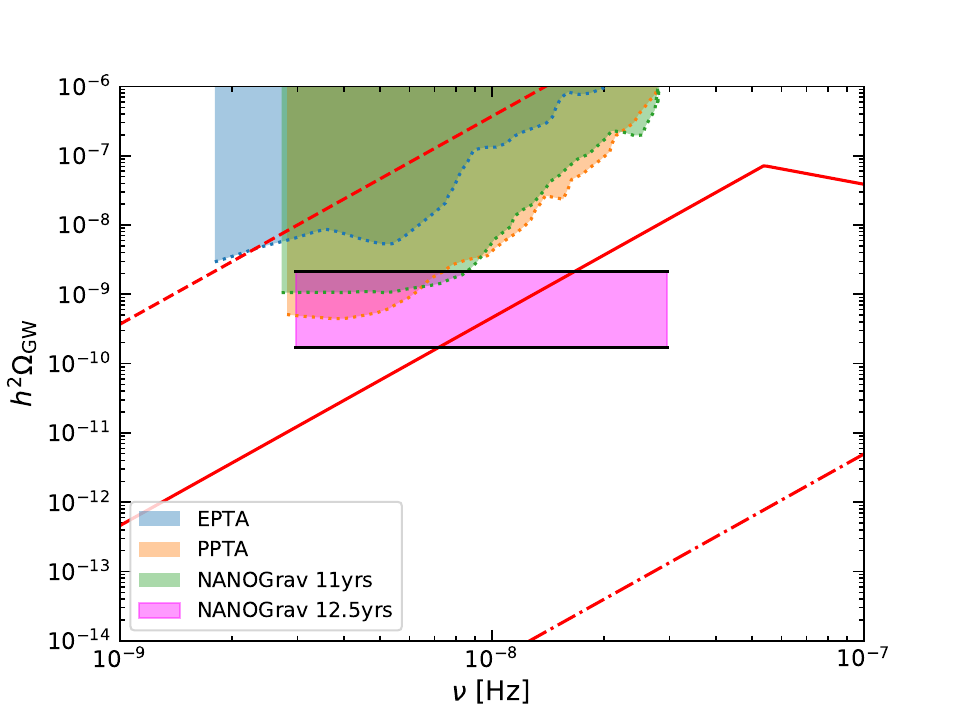}
    \includegraphics[width=75mm,angle=0]{./modeldata.pdf}
    \caption{Left plot: from Ref.~\cite{Chiang:2020aui}, the red solid line represents our prediction for the NANOGrav 12.5-year observations with $f=200$~TeV in the clockwork axion framework. Right plot: the fit to the NANOGrav 15-year dataset and IPTA-DR2 dataset in this {\it Letter}.}
    \label{fig:predcomp}
\end{figure}

In Fig.~\ref{fig:predcomp}, we plot our prediction of the nano-Hz GW spectrum for NG12 data in Fig.~10 of Ref.~\cite{Chiang:2020aui} together with the results by fitting to the recent NG15 and IPTA-DR2 datasets. As shown by this figure, as the PTA data accumulates, the nano-Hz GW observations tend to be consistent with our prediction, both in the amplitude and in the exponent of the GW spectrum.

\section{PBH formation}

In this section, we show the potential constraints from the formation of PBH~\cite{Ferrer:2018uiu}, as argued in Ref.~\cite{Gouttenoire:2023ftk}.
The PBH would have a lifetime longer than the Universe's age if it is formed with mass $M_{\rm PBH}\gtrsim 10^{15}$~g and, therefore, is considered as a DM candidate. During the cosmological evolution, the DW will oscillate and shrink in size when its scale tends to exceed 
the time scale. The plasma friction may also lead to DW shrinking but is found to be negligible. 
The shrinking of the closed DW can lead to the formation of PBHs. 
However, there are still uncertainties in such a PBH formation mechanism and thus in its relic abundance. 
The PBH formation is most significant at a later time, i.e., $t_{\rm PBH}\gtrsim t_{\rm QCD}$.
We expect that only a very small fraction of DW energy will collapse into PBHs if they form after the onset of the DW annihilation.
The DW annihilates significantly when $\sigma H\sim V_b$, where $V_b$ is the bias energy lifted by the QCD instanton. 
Thus, when $t<t_{\rm ann}$, we have  
\begin{equation}\label{eq:vblst}
    V_b\lesssim \frac{\sigma}{2t}  
    ~.
\end{equation}
Before the annihilation, the PBH mass within the time scale is
\begin{equation}
    M(t) \simeq \frac{4}{3} \pi t^3 V_{\mathrm{b}}+4 \pi t^2 \sigma \lesssim 5\pi t^2\sigma
    ~,
\end{equation}
where condition~\eqref{eq:vblst} is used in the second inequality. Then the ratio of the Schwarzschild radius to the time scale is 
\begin{equation}
    p(t)=\frac{R_{\mathrm{Sch}}(t)}{t}=\frac{2 G M(t)}{t}\lesssim 10\pi Gt\sigma
    ~.
\end{equation}
The PBH forms at the time $t_{\rm PBH}$ when $p(t)$ approaches 1 with the growing time scale~\cite{Ferrer:2018uiu}. 
We thus determine a lower limit of the time for the PBH formation
\begin{equation}\label{eq:pbhtime}
    t_{\rm PBH}\gtrsim \frac{1}{10\pi G \sigma}
    ~.
\end{equation}
We should ensure the PBHs are not formed before the annihilation of DW, i.e., $t_{\rm PBH}\gtrsim t_{\rm ann}$, so that PBH formation is negligible and the Universe is not over-closed by the PBHs. This condition can be satisfied if $t_{\rm ann}\lesssim 1/(10\pi G\sigma)$, which gives
\begin{equation}\label{eq:PBHform}
    M_{\rm pl}^2\gtrsim \frac{5\pi \sigma^2}{V_b}
    ~,
\end{equation}
where $t_{\rm ann}=\sigma/(2V_b)$ and $M_{\rm pl}=1/\sqrt{G}$. With $\sigma\simeq 8\varepsilon^{1/2}f^3$ and $V_b\sim \Lambda_{\rm QCD}^4$, we then find
a constraint on the PQ $U(1)$ symmetry-breaking scale
\begin{equation}
    f\lesssim \left( \frac{M_{\rm pl}^2\Lambda_{\rm QCD}^4}{320\pi \varepsilon} \right)^{1/6}\simeq 347.3\varepsilon^{-1/6}
    \left( \frac{\Lambda_{\rm QCD}}{330{\rm MeV}}\right)^{2/3}~{\rm TeV}
    ~.
\end{equation}
We observe that this constraint is a little weaker than the bound from DW dominating the Universe. In the case with $f\sim 200$~TeV that is favored by the PTA data, we confirm that the PBHs are indeed formed after the DW annihilation is completed.

Following~\cite{Ferrer:2018uiu}, the fraction of DM energy density in the form of PBHs after the DW annihilation is given by
\begin{equation}
    f_{\rm PBH}\equiv \Omega_{\rm PBH}/\Omega_{\rm DM} \sim p^\mathcal{N}\left( \frac{T_{\rm ann}}{T_{\rm PBH}} \right)^{3-\alpha}
    ~,
\end{equation}
where $\alpha\simeq 7$ from simulations~\cite{Kawasaki:2014sqa} and $p^{\mathcal{N}}\lesssim \mathcal{O}(1)$.
Assuming that the PBH formation takes place at $t_{\rm PBH}\simeq 1/(10\pi G\sigma)$, we have $T_{\rm PBH}/T_{\rm ann}\simeq 0.16$. Then we find
$f_{\rm PBH}\simeq 6.6\times 10^{-4}$ with $f\simeq 200$~TeV. Since $T_{\rm PBH}$ is below $1/(10\pi G\sigma)$, 
the fraction of PBHs in the DM $f_{\rm PBH}$ would be less than $\sim 10^{-4}$. 
The PBH mass is then found to be $M_{\rm PBH}\gtrsim 35M_{\odot}$. Note that the exact $f_{\rm PBH}$ and $M_{\rm PBH}$ are very sensitive to $t_{\rm PBH}$. We conclude that the Universe today would not be over-closed by the PBHs
formed after the DW annihilation.


\end{document}